\documentclass[%
 reprint,
superscriptaddress,
 aps,
prb
]{revtex4-2}

\usepackage{graphicx}
\usepackage{dcolumn}
\usepackage{bm}
\usepackage{braket}
\usepackage{ amsmath,amssymb}
\usepackage{tabularx}
\usepackage[justification=centerlast,labelfont=normal,textfont=small]{caption}
\usepackage[justification=centerlast]{subcaption}
\usepackage{hyperref}
\usepackage{gensymb}
\usepackage{cancel}
\usepackage{siunitx}
\usepackage{physics}
\usepackage[normalem]{ulem}
\hypersetup{colorlinks=true, citecolor=blue, urlcolor=blue, linkcolor=blue}

\newcommand{\be}{\begin{equation}}
\newcommand{\ee}{\end{equation}}
\newcommand{\ben}{\begin{eqnarray}}
\newcommand{\een}{\end{eqnarray}}
\newcommand{\bes}{\begin{subequations}}
\newcommand{\ees}{\end{subequations}}
\newcommand{\bF}{\begin{figure}}
\newcommand{\eF}{\end{figure}}

\usepackage{lineno}

\newcommand{\misc}{\mathrm{MISC}_{\epsilon}}
\newcommand{\proj}[1]{\left |{#1} \left \rangle \! \right \langle {#1}\right |}

\begin{document}


\title{Model-Independent Simulation Complexity of Complex Quantum Dynamics}

\author{Aiman Khan}
\email{aiman.khan@warwick.ac.uk}
\affiliation{Department of Physics, University of Warwick, Coventry, CV4 7AL, United Kingdom}

\author{David Quigley}
\affiliation{Department of Physics, University of Warwick, Coventry, CV4 7AL, United Kingdom}

\author{Max Marcus}
\affiliation{Physical \& Theoretical Chemistry Laboratory, Department of Chemistry, University of Oxford, Oxford OX1 3QZ, United Kingdom}

\author{Erling Thyrhaug}
\affiliation{Dynamical Spectroscopy, Department of Chemistry, Technical University of Munich, 85748 Garching, Germany}

\author{Animesh Datta}
\email{animesh.datta@warwick.ac.uk}
\affiliation{Department of Physics, University of Warwick, Coventry, CV4 7AL, United Kingdom}

\date{\today}

\begin{abstract}

We present a model-independent measure of dynamical complexity based on simulating complex quantum dynamics using stroboscopic Markovian dynamics. 
Tools from classical signal processing enable us to infer the Hilbert space dimension of a complex quantum system evolving under a time-independent Hamiltonian via pulsed interrogation.
We evaluate our model-independent simulation complexity (MISC) for the spin-boson model and simulated third-order pump-probe spectroscopy data for exciton transport in coupled dimers with vibrational levels.
The former provides insights into coherence and population dynamics in the two-level system while the latter reveals the dimension of the singly-excited manifold of the dimer.
Finally, we probe the complexity of excitonic transport in light harvesting 2 (LH2) and Fenna-Matthews-Olson (FMO) complexes using data from two recent nonlinear ultrafast optical spectroscopy experiments. 
For the latter we make some model-independent inferences that are commensurate with model-specific ones.
This includes estimating the fewest number of parameters needed to fit the experimental data and identifying the spatial extent, i.e., delocalization size, of quantum states participating in this complex quantum dynamics.

\end{abstract}

\maketitle

\renewcommand{\thefigure}{\Roman{figure}}
\renewcommand{\thesubfigure}{\arabic{subfigure}}

\section{Introduction}

Some of the most challenging yet exciting entities at the vanguard of our understanding in the physical sciences are complex quantum systems, ranging from molecular processes on the femtosecond timescale, such as many-body coherent dynamics in semiconductors~\cite{Cundiff2018}, exciton transport processes in photosynthetic light harvesting complexes~\cite{Chenu:2015aa}, ultrafast isomerisation in rhodopsin which is the primary photochemical event in vision~\cite{Polli2010}, to various processes in atomic, molecular, condensed matter, chemical, laser and nuclear physics~\cite{akulin2014}. The `complexity' of any quantum dynamics must depend on the dimension of the Hilbert space which the process explores.
Identifying this dimension is precisely the challenge since typical complex quantum systems involve an extended environment coupled to a finite-dimensional central core - such as vibrational levels interacting with excitons in light-harvesting complexes, resulting in, in principle, an infinite-dimensional Hilbert space. In practice, however, complex quantum dynamics often explore a small part of the environment around the central core resulting in an effective finite-dimensional Hilbert space. It is this that we seek to identify in this work.

The term `simulation complexity' was introduced in a model-dependent approach that used one-dimensional tensor networks to approximate the system-environment joint state~\cite{luchnikov2019simulation}. The bond dimension of these networks was interpreted as the square of the effective environment dimension that can simulate open system dynamics. 
In another recent work, a fluctuation-dissipation theorem for chaotic systems at high temperatures was established, linking the time-averaged fluctuation of a probe observable to the average decay rate of the test qubit by a factor that depends on the effective Hilbert space dimension of the system and environment~\cite{nation2019ergodicity}.

It has been shown that the dimensionality of the effective system-environment quantum state can be bounded in a model-independent way using tools from the theory of classical dynamical systems and classical signal processing on time series of experimental data~\cite{wolf2009assessing,strikis2019quantum,helsen2019spectral}.
This method of delays, as it is often called, computes the size of a fictitious, extended \textit{quantum} system evolving under a fixed Markovian map that reproduces the given dynamics. We refer to this dimension-based classification of dynamical complexity as \emph{model-independent simulation complexity (MISC)}.

Similar to the method of delays, the reproduction of dynamics in a model-independent manner using a fixed Markovian map in the generalized probabilistic framework has also been achieved via other tools of subspace identification~\cite{bennink2019quantum}. These methods have provided model-independent characterisations of the Hilbert space dimension of up to three engineered qubits~\cite{strikis2019quantum,bennink2019quantum,helsen2019spectral}.
Unfortunately, MISC based on the method of delays (or other subspace identification techniques) as developed in previous works cannot be applied directly to large families of experimental scenarios without accounting for and filtering out the transient effects of the time-dependent interactions. 
Prominent amongst them are linear and nonlinear spectroscopies where finite time-dependent interactions between pulses and the complex quantum systems are used to probe complex dynamics.

In this work, we develop MISC for use on time-integrated data generated from finite time-dependent interactions. This allows us to overcome the transient effects of the time-dependent pulses. 
We then define MISC with relative error $\epsilon$ as $\misc$ for noisy signals and illustrate its use in classifying simulation complexity of the Jaynes-Cummings and spin-boson models.
Finally, we evaluate $\mathrm{MISC}_{\epsilon}$ for simulated and experimental ultrafast nonlinear spectroscopy data from exciton transport in coupled dimers, and the light harvesting 2 (LH2) and Fenna-Matthews-Olson (FMO) complex respectively.
For the latter complex quantum dynamics, we make some model-independent inferences that are commensurate with model-specific inferences and electronic structure calculations. 
This suggest a role for $\misc$ in understanding challenging complex quantum systems.

\section{Quantum System Interrogated By Pulses}

Suppose the Hamiltonian governing the quantum system is
\begin{equation}\label{eq:setup}
\hat{H}(t) = \hat{H}_0 + \hat{V}(t),
\end{equation}
where $\hat{V}(t) = \eta \hat{V}\sum_{k=1}^{N} g(t-T_{k})$, $\eta$ being the strength of system-pulse coupling, $g(t)$ is the pulse envelope, $\bm{T}=\{T_1,\dots,T_N\}$ is the set of central times of the $N$ pulses, and $\hat{V}$ is the interaction operator. $\hat{H}(t)$ encompasses, amongst other experimental scenarios, linear and nonlinear spectroscopy of complex quantum systems. In all these cases, the time-integrated signal is 
\begin{equation}\label{eq:Eeq}
E(\bm{T}) = \int_{-\infty}^{\infty} dt~ _I \! \bra{\psi(t)} \hat{V}_I(t) \ket{\psi(t)}_I,
\end{equation}
where $\ket{\psi(t)}_I = e^{i\hat{H}_0 t}\ket{\psi(t)}$, $\hat{V}_I(t) = e^{i\hat{H}_0 t}\hat{V}e^{-i\hat{H}_0 t}$ are the interaction-picture state and evolution operator respectively. 

Assuming the pulses do not overlap (the semi-impulsive limit), $g(t-T_{k})g(t-T_{l}) = 0 \,\,\forall \,\,T_k \neq T_l\in \bm{T}$, and the interaction operator and system Hamiltonian do not commute $[\hat{H}_0,\hat{V}] \neq 0$, the time-integrated signal is (see Appendix \hyperref[appendix:a]{A})
     \begin{equation}\label{eq:Efirst}
       E(\bm{T}) = i\hbar \sum_{k,l=0}^{\infty} \sum_{m=0}^{k} E_{k-m,l+m+1}(\bm{T})
     \end{equation}
where $E_{kl}(\bm{T}) =~_I\!\bra{\psi^{k}(\infty)}\ket{\psi^{l}(\infty)}_I$ is the overlap of the interaction-picture asymptotic wavefunctions, defined as $\ket{\psi^{(n)}(t)}_I = \hat{U}_n(t,t_0)\ket{\psi_0}$ where
\begin{align}\label{eq:dysonmain}
  &\hat{U}_{I}(t,t_0) =  \sum_{n=0}^{\infty}\hat{U}_n(t,t_0),
 \noindent\\
   &\hat{U}_n(t,t_0) = \left(-\frac{i}{\hbar}\right)^n \int_{t_0}^{t}dt_1...\int_{t_0}^{t_{n-1}}dt_n \hat{V}_{I}(t_1)...\hat{V}_I(t_n).  \nonumber
\end{align}
To simplify the subsequent exposition, we focus on the signal as the function of a single time difference $\delta T_{\alpha} = T_{\alpha +1 } - T_{\alpha}$, in which case each term in Eq.~(\ref{eq:Efirst}) is of the form (see Appendix \hyperref[appendix:a]{A})
\begin{equation}
\label{eq:Equant}
        E_{kl}(\delta T_{\alpha}) = \sum_{\bm{\lambda}} c_{kl}(\bm{\lambda}) \chi_{abcd}(\delta T_{\alpha}) + \mathrm{constant},
\end{equation}
$\bm{\lambda} = \{\lambda_a,\lambda_b,\lambda_c,\lambda_d\}$ is a quadruplet of indices, each indexing a complete set of basis vectors $\sum_a \proj{\lambda_a}=I$ and so on, and $\chi_{abcd}(t) = \bra{\lambda_a}e^{i\hat{H}_0 t}\ket{\lambda_b}\bra{\lambda_c}e^{-i\hat{H}_0 t}\ket{\lambda_d}$ are process tensor elements corresponding to unitary evolution effected by Hamiltonian $\hat{H}_0$. The coefficients $c_{kl}(\bm{\lambda})$ are defined in Appendix \hyperref[appendix:a]{A}.
 
Using Eq.~(\ref{eq:Efirst}) and Eq.~(\ref{eq:Equant}), the time-integrated signal is, up to an additive constant, 
\begin{equation}
\label{pulseq}
    E(\delta T_{\alpha}) =    i\hbar \sum_{k,l=0}^{\infty} \sum_{m=0}^{k}  \sum_{\bm{\lambda}}   c_{k-m,l+m+1}(\bm{\lambda}) \chi_{abcd}(\delta T_{\alpha}),
 \end{equation}
a linear combination of process tensor elements that evolve via the  time-independent Hamiltonian $\hat{H}_0$ only. 
Importantly, this form allows us to sidestep the additional complexity imposed by transient effects of the time-dependent interaction Hamiltonian $\hat{V}(t)$ and instead capture the complexity of quantum dynamics induced by $\hat{H}_0$ only. 
The form of the signal in Eq.~(\ref{pulseq}) is now ready for discrete time series analysis using the method of delays to bound the complexity of quantum dynamics.
The method of delays applied directly to transient effects can lead to the identification spuriously high complexity as discussed in Appendix \ref{appendix:d}.

\emph{Method of Delays:} Consider a finite stream of data $\bm{A} \equiv A(k) \in \mathbb{R}, k \in \mathbb{N}_K \equiv \{0,1,\dots,K\}$, recorded as a discrete function of a relevant dynamical control variable indexed by $k$, typically time.
This encompasses $ E(\delta T_{\alpha})$ discussed previously (where the dynamical control variable would be the time delay between successive pulses).
A numerical value of MISC can be extracted from the data by invoking two related results from the method of delays in quantum information theory~\cite{wolf2009assessing,strikis2019quantum,helsen2019spectral}.
Firstly, given a discrete bounded time series,  there always exist an initial quantum state $\hat{\rho}_0$, a fixed generator of stroboscopic Markovian dynamics $\mathcal{P}$, and quantum observable $\hat{A}$ acting on Hilbert space of dimension $ \mathrm{rank}(M)\,+\,2$ where 
\begin{equation}
        M_{mn} = A(m+n-2); \,\,\,\,\, m,n \in\{1,\dots,K\},
\end{equation}
such that $A(k) = \mathrm{Tr}(\hat{A} \mathcal{P}^k\rho_0)\,\, \forall\,\, k\in \mathbb{N}_K.$
The $ \lceil K/2 \rceil\times\lceil K/2 \rceil$-sized matrix $M$ is referred to as the time delay (TD) matrix, giving the eponymous method~\cite{wolf2009assessing}.
Secondly, if the evolution of a given quantum state is in fact known beforehand to be given by the stroboscopic Markovian map generated by $\mathcal{Q}$ (so that $A(k) = \mathrm{Tr}[\hat{A}\mathcal{Q}^k\hat{\rho}_0]$), the inequality $\sqrt{\mathrm{rank}(M)} \leq d$ holds, where $d$ is the Hilbert space dimension of relevant dynamics. Combining both results~\cite{wolf2009assessing}, for \emph{arbitrary} dynamics,
\begin{equation}\label{eq:wpg}
        \sqrt{\mathrm{rank}(M)} \leq d \leq \mathrm{rank}(M)+2,
\end{equation}
where $d$ is interpreted as the dimension of the smallest quantum system that can simulate the complex, possibly non-Markovian, quantum dynamics using homogeneous, stroboscopic Markovian dynamics. As both the upper and lower bounds of $d$ are model-independent functions of the rank of the delay matrix, we define 
\begin{equation}
\mathrm{MISC} =  \sqrt{\mathrm{rank}(M)}
\end{equation}
as the model-independent simulation complexity (MISC).

For the TD matrix $M^E$ obtained from signal $E(\delta T_{\alpha})$ in Eq.~(\ref{pulseq}), a tighter bound corresponding to unitary evolutions~\cite{strikis2019quantum}
\begin{equation}\label{eq:misc_pp}
    \mathrm{rank} \left(M^E \right) \leq d_0^2 + d_0 - 1,
\end{equation}
follows, where $d_0$ is the dimension of the subspace of $\hat{H}_0$ on which $\hat{V}$ acts. 

\section{MISC with relative error}

In practice, $A(k)$ (or indeed any signal), is inevitably contaminated by noise. This affects the computation of MISC as a TD matrix $M$ constructed from a noisy $A(k)$, irrespective of the length or resolution of the time series, tends to have full rank. Instead, we evaluate the numerical rank of $M$ which requires, first and foremost, a meaningful delineation of the singular value spectrum of $M$ into noisy and non-noisy components. The singular value spectrum is defined by the decomposition $M = \sum_{i=1}^{ \lceil K/2 \rceil} O_1 \Sigma_i O_2,$ where $O_{1,2}$ are orthogonal matrices and $\Sigma_i$ are the singular values in descending order. In order to evaluate numerical rank, the singular values attributed to noise are set to zero
\footnote{Note that the decomposition of the time series into principal components employed here, each of which correspond to rank-$1$ TD matrices, has close resemblance to the model-free singular spectrum analysis~\cite{golyandina2001analysis}.}.

To this end, consider the reconstructed signal time series $\bm{A}_r$ formed from the TD matrix $M_r =\sum_{i=1}^{r} O_1 \Sigma_i O_2$, and define the root mean square perturbation as,
\begin{equation}\label{eq:deltar}
\Delta_r = \frac{||\bm{A}_r - \bm{A}||_2}{|| \bm{A} ||_2},
\end{equation}
where $||\cdot ||_2$ is the $2$-norm, defined as $|| \bm{A} ||_2 = \sqrt{ \sum_{k=0}^{K} |A(k)|^2  }$.
Starting with the intuition that noise will only contribute to small values in the singular value spectrum~\cite{ubaru2016fast} of the delay matrix $M$, we define the MISC with relative error $\epsilon$ as
\begin{equation}
\label{eq:misce}
\mathrm{MISC}_{\epsilon} = \sqrt{R}, ~~\mathrm{where} ~~ R = r : \Delta_{r} < \epsilon < \Delta_{r-1}
\end{equation}
for $r \in \{1,\dots,\lceil K/2 \rceil\}.$ Here, $\epsilon$ captures the relative error with which the dynamics captured by the discrete time series $A(k)$ are reproduced by the stroboscopic Markovian simulator. The choice of $\epsilon$ is determined by the signal-to-noise ratio (SNR) in $A(k)$ which sets the meaningful precision to which it is reproduced. For numerical simulations, $\epsilon$ is proportional to the precision of the numerical solver employed, whereas for experimental data it is inversely proportional to its SNR.

An important consideration in the evaluation of $\misc$ is its dependence on the length and resolution of the time series -- $K$ data points can reveal a maximum simulation complexity of $\sqrt{\lceil K/2 \rceil}$, and excessive coarse-graining of observations progressively reduces the computed $\misc$. Depending on the complex quantum system at hand there may be an insufficient amount of data, which may pose a challenge in certain experiments~\cite{strikis2019quantum}. This is indeed revealed in a single-molecule spectroscopy experiment we analyse in a later section.

\begin{figure}[h!]
    \centering
    \includegraphics[width=1.0\linewidth]{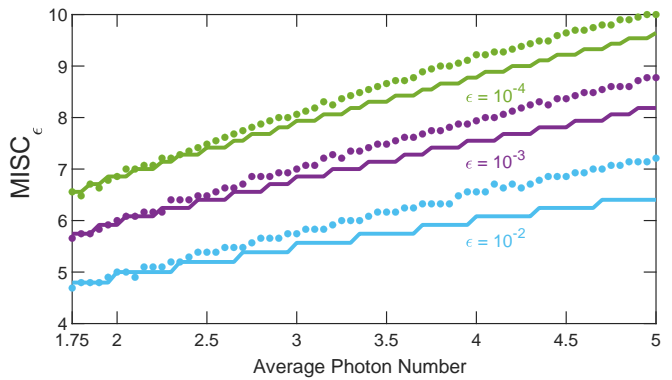}
    \caption{$\mathrm{MISC}_{\epsilon}$ (solid circle marker) and corresponding lower bounds (solid line in same colour), for time series constructed from expectation values of the population inversion operator $\langle \hat{\sigma}_z(t) \rangle$ in Eq.~(\ref{eq:13}), plotted versus the average photon number $\bar{n}$ of thermal mode for $\epsilon=10^{-4},10^{-3}$ and $10^{-2}$.}
\label{fig:jc}
\end{figure}

As expected of any reasonable measure of simulation complexity, $\misc$ also depends on the tolerance up to which we seek to reproduce the complex quantum system dynamics. As is also expected, Eq.~(\ref{eq:misce}) shows that \emph{ceteris paribus}, $\misc$ is a monotonically decreasing function of $\epsilon.$ This is illustrated for a simple theoretical model in Fig.~\ref{fig:jc}, one of the two model open quantum systems studied in the next section. 
We then turn to evaluating $\misc$ for exciton transport in photosynthetic light harvesting complexes, using (simulated) pump-probe and (experimental) two-dimensional electronic spectroscopy data.

\emph{Jaynes-Cummings (JC) Model:} The JC Hamiltonian describes linear coupling between a two-level system (TLS) (with levels notated as $\ket{0}$ and $\ket{1}$) and a single bosonic mode (for example, in a high-Q cavity), given by~\cite{jaynes1963comparison}
\begin{align}
            \hat{H} &= \frac{1}{2}\hbar\omega_0\hat{\sigma}_+\hat{\sigma}_- + \hbar\omega\hat{b}^{\dag}\hat{b} +\hbar\lambda(\hat{\sigma}_+\hat{b} + \hat{\sigma}_-\hat{b}^{\dag}),
\end{align}
where $\hat{\sigma}_+=\hat{\sigma}_-^{\dag}=\ket{1}\bra{0}$. Starting in the product initial state $ \rho(0) = \proj{1} \otimes \rho_{\mathrm{Th}}$ with
        \begin{equation}
           \rho_{\mathrm{Th}} = \frac{1}{1+\bar{n}}\sum_{n=0}^{\infty} \! \left(\frac{\bar{n}}{1+\bar{n}}\right)^n \!\!\ket{n}\bra{n},
        \end{equation}
 being the single mode thermal field at inverse temperature $\beta = 1/k_B T$ and $\bar{n} = 1/(e^{\beta\hbar\omega}-1)$ is the average photon number of the field, the expectation value of the inversion operator $\hat{\sigma}_z = \ket{1}\bra{1} - \ket{0}\bra{0}$ as a function of time is \cite{gerry2005introductory}        \begin{equation}\label{eq:13}
            \langle \hat{\sigma}_z(t) \rangle = \sum_{n=0}^{\infty} \left(\frac{\bar{n}}{1+\bar{n}}\right)^n \cos(\lambda t\sqrt{n+1})
        \end{equation}
In Fig.~\ref{fig:jc} we plot $\mathrm{MISC}_{\epsilon}$ computed using time series constructed from $\langle \hat{\sigma}_z(t)\rangle$, against average photon number. The monotonically increasing simulation complexity suggests that the number of levels with substantial participation in dynamics increases as the inverse temperature declines. A lower bound to $\mathrm{MISC}_{\epsilon}$, also plotted in Fig.~\ref{fig:jc} in corresponding colour, can be obtained analytically using a perturbation theory of singular values (See Appendix \ref{appendix:b}).

\emph{Dissipative and Dephasing Processes in the Spin-Boson Model:} Spin-boson models are a helpful prototype for understanding dissipative and dephasing dynamics in a multitude of complex quantum systems~\cite{leggett1987dynamics,fujihashi2018intramolecular}. 
Here we compute $\mathrm{MISC}_{\epsilon}$ for the spin-boson model of a TLS (with characteristic frequency $\omega_0$) coupled to a bosonic environment, described by the total Hamiltonian~\cite{leggett1987dynamics}
\begin{eqnarray}\label{eq:sbhamiltonian}
    \hat{H}_{tot} &=& \frac{1}{2}\hbar\omega_0(\hat{\psi}^{\dag}\hat{\psi} + \hat{\psi}\hat{\psi}^{\dag}) + \sum_j \left( \frac{\hat{p}_j^2}{2m_j} + \frac{m_j\omega_j^2\hat{x}_j^2}{2} \right) \nonumber \\ 
    &-& \frac{1}{2}\hat{W}\otimes\sum_j c_j\hat{x}_j +  \hat{W}^2\sum_j\frac{c_j^2}{2m_j\omega_j^2}
\end{eqnarray}
where $\hat{\psi} $ $(\hat{\psi}^{\dag})$ are fermionic annihilation (creation) operators, and $\hat{x}_j,\hat{p}_j,m_j,\omega_j$ are coordinate, momentum, mass and frequency of the $j$-th bath oscillator. The system-bath interaction, $\hat{H}_I = -\hat{W}\sum c_j\hat{x}_j$, is characterised by the dimensionless TLS operator $\hat{W}$ as well as the coupling strengths of the $j$-th oscillator $c_j$. In general,
\begin{equation}
    \hat{W} = W_1 \,.\, (\hat{\psi} \,+\, \hat{\psi}^{\dag} ) + W_2 \,.\, (\hat{\psi}^{\dag}\hat{\psi} \,+\, \hat{\psi}\hat{\psi}^{\dag}),
\end{equation}
for real numbers $W_1,W_2$, where the inelastic portion $\hat{\psi}+\hat{\psi}^{\dag}$ induces both dissipation and dephasing relaxations, and the elastic portion $\hat{\psi}^{\dag}\hat{\psi} \,+\, \hat{\psi}\hat{\psi}^{\dag}$ induces pure dephasing in the TLS. 

The influence of the bath is captured by a Drude-Lorentz-like power spectral density~\cite{ishizaki2020prerequisites}
\begin{equation}\label{eq:sd}
    J(\omega) = \frac{\hbar\zeta}{\pi\omega_0}\frac{\gamma^2\,\,\gamma_+\gamma_-\omega}{(\omega^2+\gamma_+^2)(\omega^2+\gamma_-^2)},
\end{equation}
where $\gamma_{\pm} = \gamma \pm i\delta$ ($\delta$ being a small parameter that is set to $\gamma/4$ in calculations) is the inverse correlation time of the bath noise and $\zeta$ is the strength of the system-bath coupling. The form of the spectral density Eq.~(\ref{eq:sd}) fixes problematic transient behaviour of the bath response function derived using the traditional Drude-Lorentz form~\cite{ishizaki2020prerequisites}. Further, our results for $\misc$ do not differ significantly for the standard Drude-Lorentz density. The dynamics itself is solved using the reduced hierarchy equations~\cite{ishizaki2005quantum,ishizaki2020prerequisites} (See Appendix \ref{appendix:e}).

\begin{figure}
     \centering
     \begin{subfigure}[h]{0.5\textwidth}
         \centering
         \includegraphics[width=\textwidth]{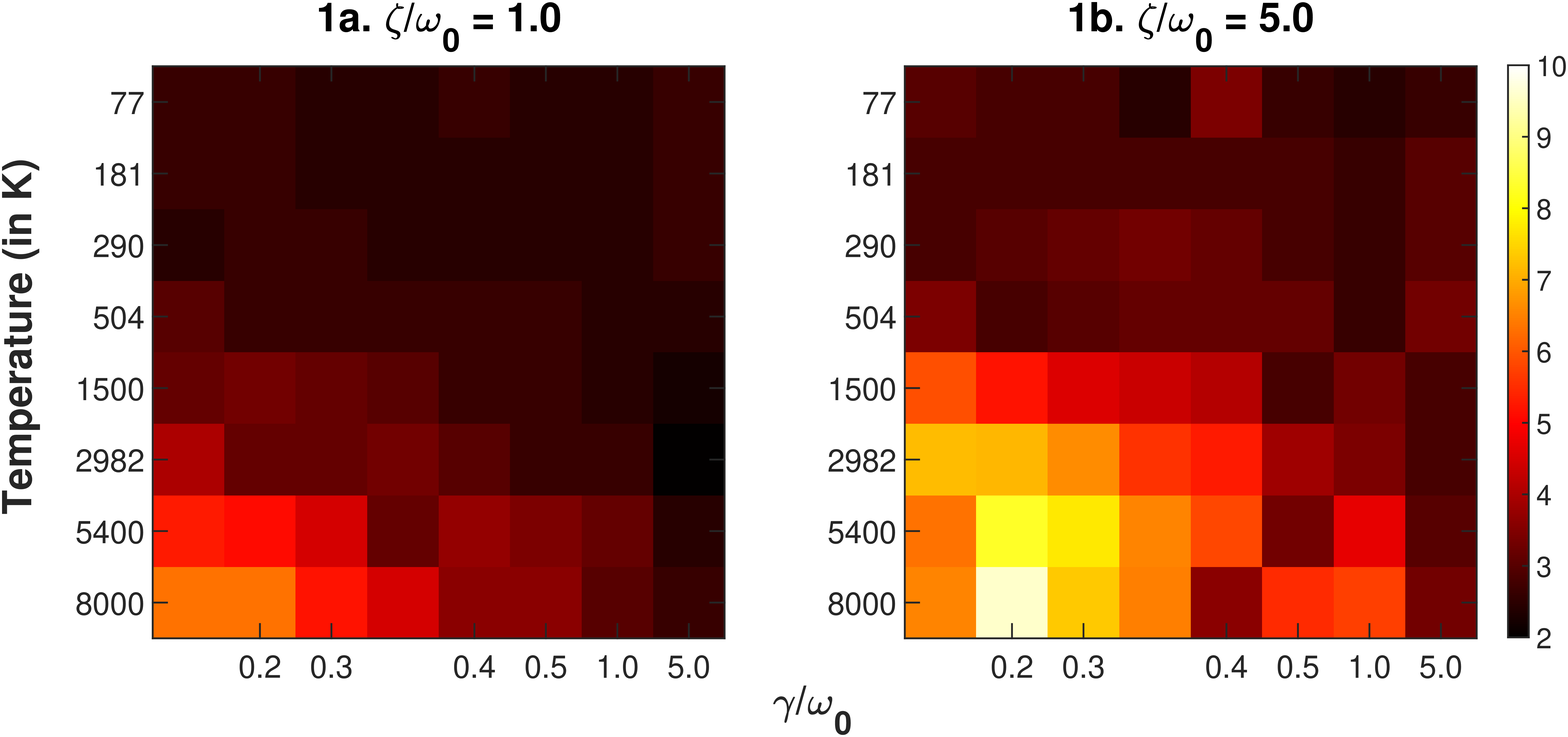}
     \end{subfigure}
     \hfill
     \begin{subfigure}[h]{0.5\textwidth}
         \centering
         \includegraphics[width=\textwidth]{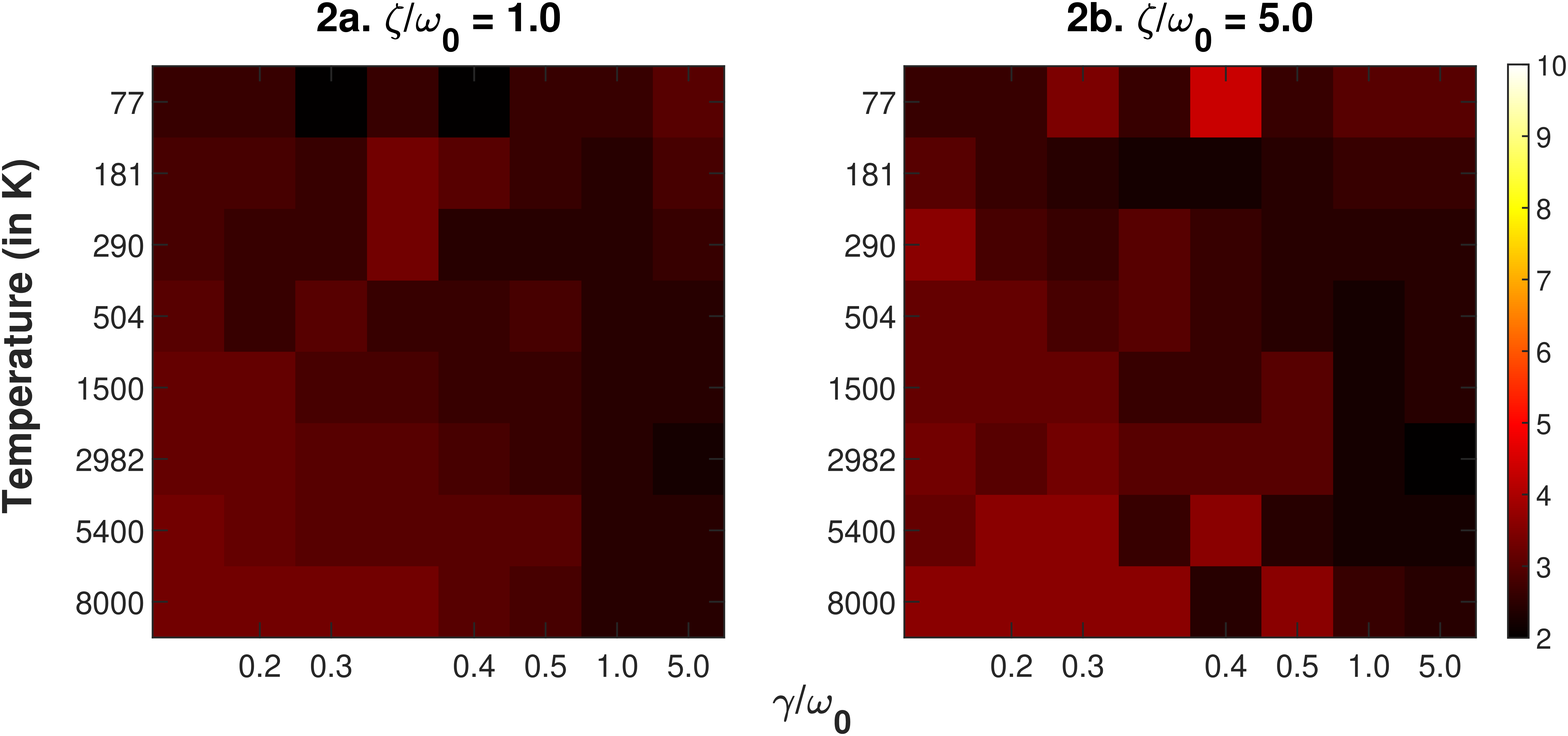}
     \end{subfigure}
        \caption{$\misc$ computed from time series of expectation value of system observables describing energy dissipation/coherence dynamics in the spin-boson model. Tolerance parameter is set to $\epsilon=10^{-4}$ . 
        (1) \emph{Inelastic dissipation process}: Time series constructed out of $\hat{\sigma}_z(t)$. (a) $\zeta/\omega_0 = 1.0$ corresponds to weak coupling between TLS and bosonic environment, and (b)  $\zeta/\omega_0 = 5.0$ corresponds to strong system-bath coupling.  
        (2) \emph{Elastic dephasing process}: Time series constructed from dynamical evolution of $\hat{\sigma}_x(t)$. (a) $\zeta/\omega_0 = 1.0$ corresponds to weak coupling between TLS and bosonic environment. (b)  $\zeta/\omega_0 = 5.0$ corresponds to strong system-bath coupling.} 
        \label{fig:sbheat}
\end{figure}

$\misc$ for both the energy relaxation induced by inelastic interaction ($W_1=1, W_2=0$) and coherence dynamics induced by elastic interaction ($W_1=0, W_2=1$)  are presented as heat maps (for a range of system-bath parameters) in Fig.~\ref{fig:sbheat}. The choices of initial states used to solve the corresponding spin-boson dynamics are detailed in Appendix \ref{appendix:e}. The heat maps indicate that the simulation complexity increases with both increasing bath correlation times (1/$\gamma$) as well as system-bath coupling strength ($\zeta$), for both population and coherence dynamics of the TLS. In the limit of high temperature and fast dynamics, $\misc$ tends to the expected Markovian value of two for a TLS. The main point of difference between the two processes is their contrasting trends with increasing temperature - while the complexity increases with temperature for inelastic population dynamics, there is no such marked increase in complexity for elastic coherence dynamics.

This behaviour of the simulation complexity can be explicitly understood through the signal time series that is used to compute $\misc$ at each point on the grids in Fig.~\ref{fig:sbheat} (which are included in  Appendix \ref{appendix:e} for both inelastic and elastic interaction Hamiltonians). For the dissipation process (Fig.~\ref{fig:sbheat} 1a and 1b), as the temperature increases the amplitude of oscillations in the populations increases resulting in higher $\misc$, negating, in some measure, the effects of faster equilibration. In contrast, expected damped coherence oscillations are observed for the coherence dynamics, which tend to a non-oscillatory decay as the temperature increases. However, the spectral content in the limit of high temperature for small values of $\gamma/\omega_0$ are still evident, thus explaining the almost constant $\misc$ for the left side of the grids in Fig.~\ref{fig:sbheat} 2a and 2b. Interestingly, $\misc$ for slow dynamics and small temperatures (top left edge of the grids in Fig.~\ref{fig:sbheat} 2a and 2b) is the same as the Markovian limit (bottom right of grids), even though the dynamics look very different.

\section{$\misc$ for Simulated \& Experimental Nonlinear Spectroscopy Data}

We next evaluate the model-independent simulation complexity with relative error $\epsilon$ -- $\misc$ -- for excitonic transport in photosynthetic light harvesting complexes~\cite{Chenu:2015aa}. These dynamics are driven by strong exciton-phonon interactions which render them complex quantum systems, and are most fruitfully studied using ultrafast nonlinear spectroscopy~\cite{Chenu:2015aa,hildner2013quantum,thyrhaug2018identification,schlau2012elucidation,westenhoff2012coherent,yuen2014ultrafast} for which the recorded signal is of the form obtained in Eq.~(\ref{pulseq}). Therefore, $\misc$ for these complex systems can be obtained precisely as for the theoretical open systems just analysed, except that the TD matrices are constructed using the time-integrated signal $E(\delta T_{\alpha})$ recorded as a function of $\delta T_{\alpha}$ and that the tighter lower bound of Eq.~(\ref{eq:misc_pp}) holds.

\emph{Simulated pump-probe spectroscopy of coupled dimer:} The strong exciton-phonon interaction in photosynthetic light harvesting complexes is often modelled using the Frenkel-Holstein Hamiltonian $ \hat{H}_{FH} = \hat{H}_{\mathrm{exc}} + \hat{H}_{\mathrm{ph}} + \hat{H}_{\mathrm{exc-ph}} $ where the successive terms denote the exciton, phonon, and interaction Hamiltonians. For a dimer
\begin{equation}\label{eq:fh1}
    \hat{H}_{\mathrm{exc}} = \sum_{i=1,2}\varepsilon_{i}\hat{c}_i^{\dag}\hat{c}_i + \kappa(\hat{c}_1^{\dag}\hat{c}_2 + \hat{c}_1^{\dag}\hat{c}_2)
\end{equation}
\begin{equation}\label{eq:fh2}
  \hat{H}_{\textrm{ph}} = \sum_{i=1,2}\hbar\omega_i(\hat{d}_i^{\dag}\hat{d}_i + 1/2) ,
\end{equation}
\begin{equation}\label{eq:fh3}
\hat{H}_{\textrm{exc-ph}} = -\sum_{i=1,2}\hbar\omega_ig_i\hat{c}_i^{\dag}\hat{c}_i(\hat{d}_i^{\dag} + \hat{d}_i),
 \end{equation}
 where $\varepsilon_i$ are the on-site energies, $\kappa$ is the strength of dipole-dipole coupling between the sites 1 and 2, $\{\hat{c}_i^{\dag},\hat{c}_i\}$ are the creation-annihilation operator pair for the two sites, $\{\hat{d}_i^{\dag},\hat{d}_i\}$ are the phonon creation-annihilation operator pair for vibrational bath coupled to each site, $\omega_i$ are the phonon frequencies for the vibrational ladders and $g_i$ are the individual strengths of the phonon-exciton coupling. 

Table \ref{tab:PPsim} displays the $\misc$ evaluated using the simulated signal generated from a sequence of four distinct pump-probe experiments~\cite{marcus2019towards} used to investigate singly-excited manifold (SEM) dynamics of the dimer, for a varying number of phonons in the vibrational bath (See Appendix~\ref{appendix:c}). Each component signal is aggregated over 300 molecules with varying dipole orientations corresponding to individual sites but fixed relative orientation of $40^{\circ}$. In order to simultaneously take into account the four distinct signals that are recorded, we construct TD tensors (in place of TD matrices) whose spanned space will be bound by the same inequality as Eq.~(\ref{eq:wpg}). The simulation complexity is then set to be the maximum of $\misc$ corresponding to all possible combinations of the signal components.
As this is a simulation, the precise dimension of the complex quantum system's Hilbert space is known. $\misc$ is expectedly found to be always less than the Hilbert space dimension of the dimer SEM which is probed by the pump-probe experiment as presented in Table~\ref{tab:PPsim}.
Furthermore, lower values of $\epsilon$ lead to larger complexity.

\begin{table}[h]
    \centering
    \begin{tabular}{| c | S[table-format=3.2] | S[table-format=3.2] | c |}
    \hline
    
         \# of Phonons  & $\mathrm{MISC}_{\epsilon=10^{-1}}$ & $\mathrm{MISC}_{\epsilon=10^{-4}}$ & SEM dimension\\
         \hline
         0\,\,\,\,\,\,  & 1.41 & 1.73 & 2 \\
         \hline
         1\,\,\,\,\,\, & 1.73 & 3.74 & 4 \\
         \hline
         2\,\,\,\,\,\, & 2.82 & 6.40 & 9 \\
         \hline
         3\,\,\,\,\,\, & 8.48 & 13.30 & 16\\
         \hline
         4\,\,\,\,\,\, & 11.74 & 21.61 & 25   \\
         \hline
    \end{tabular}
    \caption{$\misc$ for $\epsilon$ = $10^{-1}$ and $10^{-4}$ calculated from TD tensors constructed out of aggregate signals corresponding to 2800 time delays, chosen uniformly between 0.5 ps and 6.098 ps for pump-probe spectroscopy of dimer molecules. The dimer parameters are taken to be those of Allophycocyanin (APC) molecule (See Appendix~\ref{appendix:c}).}
    \label{tab:PPsim}
\end{table}

\emph{Experimental single-molecule pump-probe spectroscopy of LH2 complex:} We evaluate $\misc$ for excitonic transport in single LH2 complexes using experimental data resulting from the excitation of single LH2 molecules with two phase-coherent ultrafast pulses~\cite{hildner2013quantum}. The experiment was designed to explore quantum coherent population transfer over varying pathways in single LH2 complexes. 
The time series, constructed from 17 data points, yields a full-rank TD matrix for $\epsilon=10^{-1}$, meaning $\misc =\sqrt{9} =3.0$. This suggests that more experimental data is required to meaningfully quantify the simulation complexity of excitonic transport in LH2 complexes.

\begin{figure}[t!]
        \includegraphics[width=0.48\textwidth]{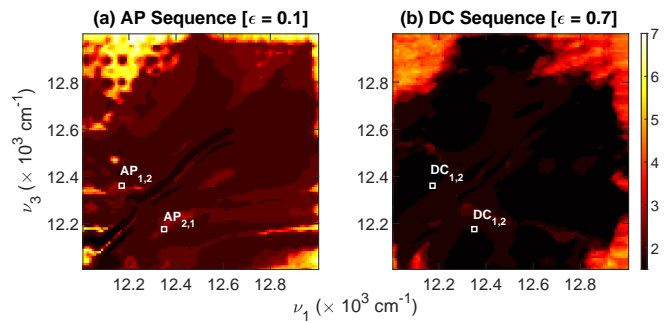}
        \label{fig:y equals x}
        \caption{Heat maps of $\misc$ for (a) AP ($\epsilon = 0.1$) and (b) DC ($\epsilon = 0.7$) pulse sequences for excitonic transport in the FMO complex using experimental data from polarization-controlled 2D electronic spectroscopy~\cite{thyrhaug2018identification}. Each maps is composed of a $100\times 100$ grid of ($\nu_1,\nu_3$) points.
         Each point is generated from the time series of the complete rephasing signal $\tilde{E}^{(3)}(\nu_1,t_2,\nu_3)$, where $\nu_1$ and $\nu_3$ are proportional to the excitation and detection energies, and the complex emitted signal field is recorded (a) over 2.4 ps for the AP sequences, and (b) over 2.9 ps for DC sequences, each sampled every 20 fs.  The cross-peaks for both pulse sequences are marked with white boxes on the map.}
        \label{fig:exheatmap}
\end{figure}

\vfill
\emph{Experimental 2D electronic spectroscopy of FMO complex:} 
Finally, we evaluate $\misc$ for excitonic transport in the FMO complex using experimental data from polarization-controlled 2D electronic spectroscopy~\cite{thyrhaug2018identification}. The original experiment uses two distinct configurations of polarizations of the pulses -- a sequence of four all-parallel (AP) pulses ($\langle 0\degree,0\degree,0\degree,0\degree\rangle$) and the double-crossed (DC) sequences of pulses ($\langle 45\degree,-45\degree,90\degree,0\degree\rangle$). 
The experiment is designed to study both the short-lived excitonic coherence as well as the long-lived vibronic coherence of FMO dynamics at 77K, recorded over several picoseconds. The spectra is generated as a sequence (for different population times $t_2$) of 2D plots of the complex emitted field, as a function of the excitation and detection energies (denoted by the wavenumbers $\nu_1$ and $\nu_3$ here). The real and imaginary components of the emitted field are used to construct the time series and the evaluated $\misc$ are presented in Fig.~\ref{fig:exheatmap}.

Some observations are in order. Firstly, the choice of different $\epsilon$ for the AP and DC data is motivated by the different SNRs in the two experiments. DC experiments have lower noise~\cite{Bukart2020}, but also detect smaller signals resulting in a lower SNR compared to the AP experiments. 
Indeed, $\misc$ for varying $\epsilon$ for the DC pulse sequence helps reveal the noise floor in the data, providing $\epsilon$ its physical interpretation (See Appendix \ref{appendix:dcmisc}). 
Secondly, the AP $\misc$ map has a diagonal valley of low simulation complexity which subsumes the vicinities of the cross-peaks linking the two lowest energy excitons where particularly prominent oscillations were noted in the original study~\cite{thyrhaug2018identification}, specifically the AP$_{2,1}$ peak that corresponds exclusively to ground state vibrations. 
Thirdly, the AP data reveals slightly higher complexity than the DC data as shown by the former's lighter shade. This is commensurate with the design of the DC experiment which suppresses some of the Liouville pathways present in the AP data to reveal certain weaker ones.
Finally, a $\misc$ of 2 across most of the DC map shows that the intermolecular excitonic coherence effectively explores a two-dimensional Hilbert space, while the intra-molecular vibrational motion that dominates the AP data involves no more than 3 Hilbert space dimensions. 
In conjunction with fact that the FMO is constituted of identical bacteriochlorophyll a molecules, this suggests that the excitonic dynamics is fairly localised spatially, in qualitative agreement with electronic structure calculations~\cite{Adolphs2006}. 

The evaluated simulation complexity, rounded off to $\lceil\misc\rceil$, corresponds to the dimension of the smallest quantum system that can possibly reproduce measured dynamics of complex systems with relative error $\epsilon$. This allows us to estimate the fewest number of parameters needed to reproduce the measured signal. To describe unambiguously stroboscopic Markovian dynamics on a Hilbert space of dimension $d$, one would require $3d^2 +d -3$ real parameters to define the initial quantum state, the map generating stroboscopic Markovian dynamics as well the Hermitian observable whose expectation is used to reproduce the signal time series. Using the computed $\lceil\misc\rceil$ for the cross peaks (which is $3$ for both the AP cross peaks in Fig.~\ref{fig:exheatmap}(a) and DC cross peaks in Fig.~\ref{fig:exheatmap}(b)), our framework predicts a minimum of 27 real parameters are required for both cross peaks of the AP and DC sequences. This compares with 24 parameters for AP cross peaks and 20 parameters for DC cross peaks used in the original study~\cite{thyrhaug2018identification}.

\section{Conclusions and Discussion}

We have developed a model-independent framework to quantify  dynamical complexity of quantum systems via $\misc$ that can be applied directly to experimental data stemming from experiments performed on complex systems. The model-independence offers unambiguous interpretation in terms of the minimum number of parameters needed to simulate this experimental data using stroboscopic Markovian dynamics upto a desired relative error $\epsilon$. We have illustrated this through numerically generated data from simple theoretical models as well as simulated and experimental data from non-linear spectroscopy experiments.

The generality and simplicity of our method makes its applicability universal. The full potential of our work can only be realised by applying it to data from future experiments on complex systems. The model-independent inferences can then be compared to model- and system-specific ones to identify the universal features of complex quantum systems, furthering our understanding of the diverse processes that dominate their dynamics. 
A particularly relevant instance is determining the extent of certain complex quantum dynamics in physical space. 
As many of these systems, such as pigment-protein complexes, are constituted of simple molecular units, the Hilbert space dimension can provide an estimate of the number of units, and thus the spatial extent of eigenstates participating in particular dynamics, aiding complicated electronic structure calculations~\cite{olbrich2011theory}.

\section{Acknowledgements}

We are grateful to D. Zigmantas and R. Hildner for sharing their experimental data and expertise. We thank E. Bittner, S. Huelga,  A. Ishizaki, G. Knee, J. Lim, K. Modi and M. Plenio for illuminating discussions over the course of this work. AK was supported by a Chancellor's scholarship from the University of Warwick and AD by an EPSRC fellowship (EP/K04057X/2). 
We acknowledge the use of the Scientific Computing Research Technology Platform of the University of Warwick,
 and Athena at HPC Midlands+ funded by the EPSRC (EP/P020232/1).

\bibliography{ReferencesComDimWit.bib}

\vfill

\appendix
\renewcommand\thefigure{\thesection.\arabic{figure}} 
\setcounter{figure}{0}

\begin{widetext}
\section{Deriving the form of the time-integrated signal $E(\delta T_{\alpha})$}\label{appendix:a}

To show that the Eqs.~(\ref{eq:Efirst}) and (\ref{eq:Equant}) (and hence the final form in Eq. (\ref{pulseq})) hold, it is simplest to work in the interaction picture. To this end, we define $\ket{\psi(t)}_I = e^{i\hat{H}_0 t/\hbar}\ket{\psi(t)}_S$ and $\hat{O}_I(t) = e^{i\hat{H}_0 t/\hbar}\hat{O}_S(t)e^{-i\hat{H}_0 t/\hbar}$, where the subscripts $S$ and $I$ denote state vectors and operators in the Schr\"{o}dinger and interaction pictures respectively. The interaction picture time evolution operator, defined by the equation $\ket{\psi(t)}_I = \hat{U}_I(t,t_0)\ket{\psi(t_0)}_I$, is given by the Dyson series
\begin{equation}\label{eq:dyson}
\hat{U}_{I}(t,t_0) =  \sum_{n=0}^{\infty}\hat{U}_n(t,t_0),~~~~\hat{U}_n(t,t_0) = \left(-\frac{i}{\hbar}\right)^n \int_{t_0}^{t}dt_1...\int_{t_0}^{t_{n-1}}dt_n \hat{V}_{I}(t_1)...\hat{V}_I(t_n).
\end{equation}
The initial time $t_0$ must be chosen so that 
\begin{equation}
    g(s) = 0,~~ -\infty<s<t_0<\mathrm{min}(\mathbf{T}) = T_1
\end{equation}
so that all system-pulse interactions happen after $t_0$. Of course, the simplest choice is to set $t_0 = -\infty$, which is what we assume henceforth.
The initial state $\ket{\psi_0}$ is assumed to be a ground state eigenfunction of the Hamiltonian $\hat{H}_0$, so that the initial Schr\"{o}dinger and interaction picture wavefunctions coincide, $\ket{\psi(t_0)}_S=\ket{\psi(t_0)}_I=\ket{\psi_0}$. The structure of the Dyson series in Eq.~(\ref{eq:dyson}) leads to the identity 
\begin{equation}\label{eq:id}
    \partial_{t}[\hat{U}_n(t,t_0) ] = \left(\frac{-i}{\hbar}\right)\hat{V}_I(t)\hat{U}_{n-1}(t,t_0).
\end{equation}
Abbreviating $\ket{\psi^{(n)}(t)} = \hat{U}_n(t,t_0)\ket{\psi_0}$, the time-integrated signal defined in Eq.~(\ref{eq:Eeq}) can be expanded as the sum
\begin{equation}\label{eq:Edeltaa}
   E(\bm{T}) = \int_{-\infty}^{\infty}dt~\sum_{k,l=0}^{\infty}~ _I\!\bra{\psi^{(k)}(t)} \hat{V}_I(t) \ket{\psi^{(l)}(t)}_I.
\end{equation}
Using the identity in Eq.~(\ref{eq:id}) and integrating by parts, we arrive at
\begin{align}\label{eq:recursion}
\int_{-\infty}^{\infty}dt& _I\!\bra{\psi^{(k)}(t)} \hat{V}_I(t) \ket{\psi^{(l)}(t)}_I  = i\hbar \int_{-\infty}^{\infty}dt~  _I\!\bra{\psi^{(k)}(t)}\left(\partial_t \ket{\psi^{(l+1)}(t)}_I\right)\nonumber\noindent\\
&=\, i\hbar~ _I\!\bra{\psi^{(k)}(\infty)}\ket{\psi^{(l+1)}(\infty)}_I  - \cancelto{0}{i\hbar~ _I\!\bra{\psi^{(k)}(-\infty)}\ket{\psi^{(l+1)}(-\infty)}_I} + \int_{-\infty}^{\infty}dt ~ _I\!\bra{\psi^{(k-1)}(t)}\hat{V}_I(t)\ket{\psi^{(l+1)}(t)}_I\noindent\nonumber\\
&= i\hbar~ _I\!\bra{\psi^{(k)}(\infty)}\ket{\psi^{(l+1)}(\infty)}_I  +  \int_{-\infty}^{\infty}dt ~ _I\!\bra{\psi^{(k-1)}(t)}\hat{V}_I(t)\ket{\psi^{(l+1)}(t)}_I.
\end{align}
In the second step, we have used the fact that the perturbative wavefunction $\ket{\psi^{(l)}(-\infty)} = 0 ~\forall~ l\in\mathbb{N}$.
Further, the recursion allows us to successively reduce the form of Eq.~(\ref{eq:recursion}) to the following sum of overlaps of asymptotic perturbative component wavefunctions
\begin{align}
 \int_{-\infty}^{\infty}dt _I\!\bra{\psi^{(k)}(t)} \hat{V}_I(t) \ket{\psi^{(l)}(t)}_I  &= i\hbar~ _I\!\bra{\psi^{(k)}(\infty)}\ket{\psi^{(l+1)}(\infty)}_I  + i\hbar~ _I\!\bra{\psi^{(k-1)}(\infty)}\ket{\psi^{(l+2)}(\infty)}_I  \nonumber\noindent\\
 &+ \int_{-\infty}^{\infty}dt~  _I\!\bra{\psi^{(k-2)}(t)}\hat{V}_I(t)\ket{\psi^{(l+2)}(t)}_I \noindent\\
 &= i\hbar \sum_{m=0}^{k-1}~ _I\!\bra{\psi^{(k-m)}(\infty)}\ket{\psi^{(l+m+1)}(\infty)}_I + \int_{-\infty}^{\infty}dt~  _I\!\bra{\psi^{(0)}(t)}\hat{V}_I(t)\ket{\psi^{(k+l)}(t)}_I, \nonumber \\
  &= i\hbar \sum_{m=0}^{k}~ _I\!\bra{\psi^{(k-m)}(\infty)}\ket{\psi^{(l+m+1)}(\infty)}_I,
\end{align}
and we have used $ \bra{\psi^{(0)}} \ket{\psi^{(k+l+1)}(-\infty)}=0.$ The time-integrated signal  $E(\bm{T})$ is then
\begin{equation}
E(\bm{T})= i\hbar \sum_{k,l=0}^{\infty} \sum_{m=0}^{k}~ _I\!\bra{\psi^{(k-m)}(\infty)}\ket{\psi^{(l+m+1)}(\infty)}_I,
\end{equation}
which is Eq.~(\ref{eq:Efirst}).

Assuming that the pulse envelopes are sharp enough to be faithfully approximated by Dirac delta functions, such that $g(s-T_{i}) = \delta(s-T_{i})$, each component $E_{kl}(\bm{T}) =~_I\!\bra{\psi^{k}(\infty)}\ket{\psi^{l}(\infty)}_I$ 
 in the above summation is then (upto a factor of $\left(i\eta/\hbar\right)^{k+l}$)
\be
\label{eq:ekleqn}
    E_{kl}(\bm{T}) =  \sum_{\bm{\tau},\bm{\tau'}}
    \bra{\psi_0} e^{iH_0\tau'_1}\hat{V}e^{iH_0(\tau'_2-\tau'_1)}\hat{V}  \dots e^{-iH_0\tau'_k} e^{iH_0 \tau_l}\hat{V} e^{-iH_0(\tau_{l}-\tau_{l-1})} ... Ve^{-iH_0(\tau_2-\tau_1)}\hat{V}e^{-iH_0 \tau_1}\ket{\psi_0},
\ee
where $\bm{\tau}' \equiv \{\tau'_1\leq\dots\leq\tau'_k \}, \bm{\tau} \equiv \{ \tau_1\leq\dots\leq\tau_l \}$ are ascending sequences of pulse central times such that $\tau_i,\tau_i' \in \bm{T}$.

Although the signal depends on the $N$ different pulse times $\bm{T}$, for simplicity we now focus on only one time difference $\delta T_{\alpha} = T_{\alpha +1 } - T_{\alpha}$. The ascending sequences then allow
us to partition the signal into contributions before and after $T_{\alpha}$ as (upto a factor of $\left(i\eta/\hbar\right)^{k+l}$)
\ben
\label{eq:Ekl}
&&      E_{kl(\delta T_\alpha)} =  \left( \sum_{\bm{\tau}':\tau'_k < T_{\alpha}} \sum_{\bm{\tau}:\tau_l < T_{\alpha}} \right)\bra{\psi_0}e^{iH_0 \tau'_1}\hat{V} e^{iH_0(\tau'_2-\tau'_1) }\dots e^{-i\hat{H}_0(\tau'_k-\tau_l)}\hat{V}e^{-iH_0(\tau_l-\tau_{l-1})}\dots \hat{V}e^{-i\hat{H}_0\tau_1}\ket{\psi_0} \noindent\\
   & &+ \sum_{\substack{\bm{\lambda}}} \left( \sum_{\bm{\tau}':\tau'_k \geq T_{\alpha}} \sum_{\bm{\tau}} + \sum_{\bm{\tau}'} \sum_{\bm{\tau}:\tau_l \geq T_{\alpha}}  \right) \bra{\psi_0} e^{iH_0 \tau'_1}\hat{V}\dots\hat{V}\ket{\lambda_a} \bra{\lambda_b}\hat{V}\dots\hat{V}\ket{\lambda_c} \bra{\lambda_d}\hat{V}\dots\hat{V}e^{-iH_0 \tau_1}\ket{\psi_0} \chi_{abcd}(\delta T_\alpha) \nonumber
\een
where $\chi_{abcd}(t) = \bra{\lambda_a}e^{iH_0 t}\ket{\lambda_b}\bra{\lambda_c}e^{-iH_0 t}\ket{\lambda_d}$ are the process tensor elements~\cite{yuen2014ultrafast} corresponding to unitary evolution generated by $\hat{H}_0.$ This form is obtained by inserting resolutions of the identity, $I = \sum_a \proj{\lambda_a}$ in terms of the complete set of bases $\ket{\lambda_a}$ for the Hilbert space in which $\hat{H}_0$ resides, four times denoted by $\bm{\lambda} \equiv \{\lambda_a,\lambda_b,\lambda_c,\lambda_d\}$ around $T_{\alpha}$ in Eq.~(\ref{eq:ekleqn}).

The first part of Eq.~(\ref{eq:Ekl}) is independent of $\delta T_{\alpha}$, corresponding to the action of pulses interacting with the system $T_{\alpha}$ and hence is taken to be a constant for remainder of our derivation. The form of the quantity
\begin{equation}\label{eq:ckl}
    c_{kl}(\bm{\lambda}) = \left( \sum_{\bm{\tau}':\tau'_k \geq T_{\alpha}} \sum_{\bm{\tau}} + \sum_{\bm{\tau}'} \sum_{\bm{\tau}:\tau_l \geq T_{\alpha}}  \right)
    \!\! \bra{\psi_0}e^{iH_0 \tau'_1}\hat{V}\dots \hat{V}\ket{\lambda_a} \bra{\lambda_b}\hat{V}\dots\hat{V}\ket{\lambda_c} \bra{\lambda_d}\hat{V}\dots\hat{V}e^{-iH_0 \tau_1}\ket{\psi_0}
\end{equation}
shows that it is nonzero \textit{only} on the subspace of the Hilbert space of $\hat{H}_0$ spanned by $\hat{V}$. This leads to Eq.~(\ref{eq:Equant}).

\section{$\misc$ for transient effects in dynamics for time-dependent Hamiltonians}\label{appendix:d}

The time series corresponding to expectation value of any Hermitian observable $\hat{O}$,
\be
o(t) = ~ _I \! \bra{\psi(t)} \hat{O}_I(t) \ket{\psi(t)}_I
\ee
constructed with respect to the experimental time for time-dependent Hamiltonian regimes as in Eq.~(\ref{eq:setup}) can also be studied using the method of delays. In this case, the calculated $\misc$ may suggest a spuriously high complexity as $o(t)$ includes the effects of transient dynamics caused by the time-dependent interaction Hamiltonian $\hat{V}(t)$, in addition to the dynamics associated with the time-independent Hamiltonian $\hat{H}_0.$  We illustrate this with two simple examples for which $\misc=2$ for $\epsilon=0$. 

\begin{figure}

    \centering
    \includegraphics[width=0.8\linewidth]{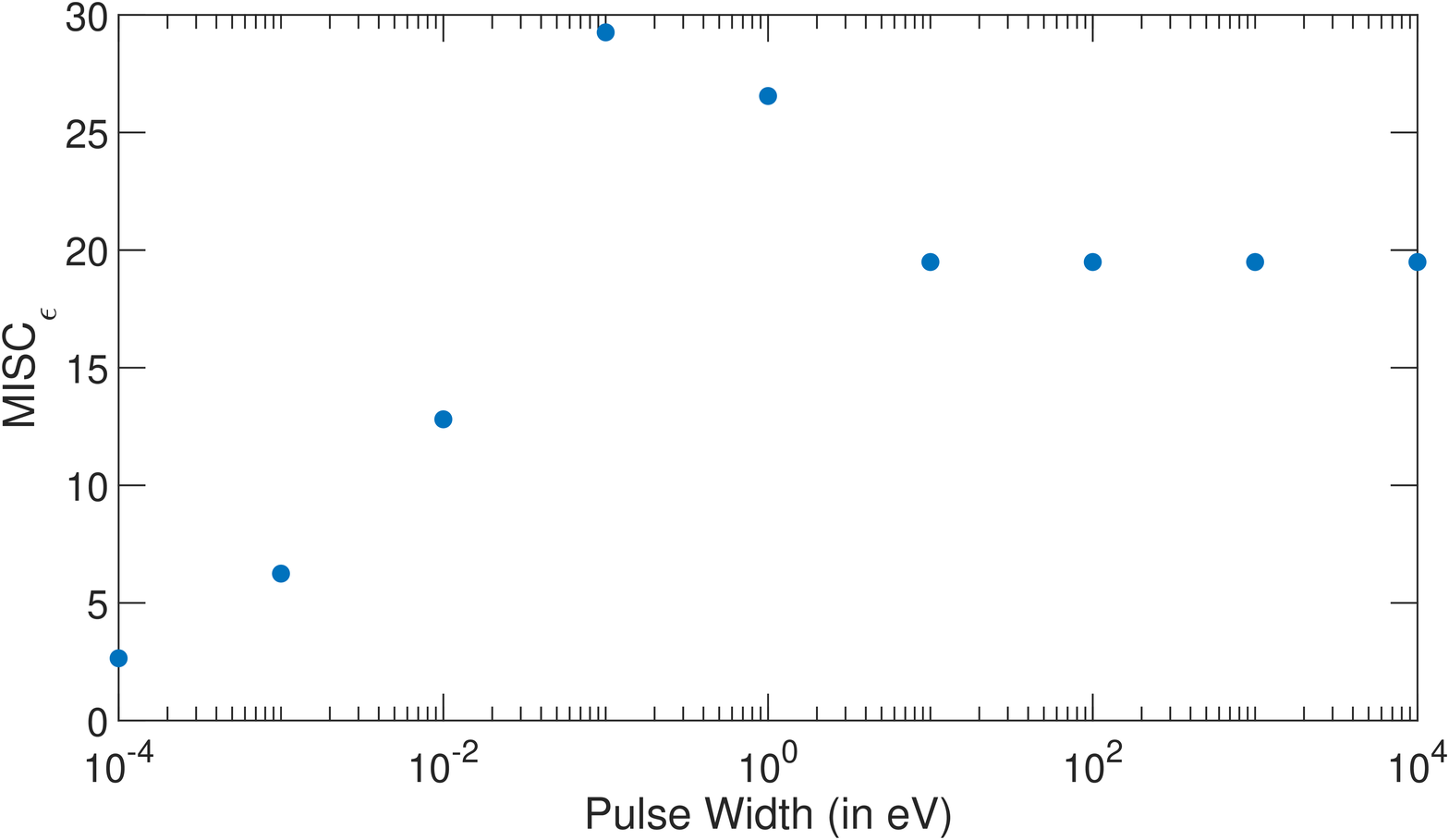}
    \caption{Numerically computed simulation complexity $\misc$ evaluated using time series (composed of 10000 values uniformly sampled in time interval extending from $0$\,\,ps to $5$\,\,ps) derived from the expectation values of the population inversion operator in Eq.~(\ref{eq:sigmaztransient}), versus the width of the pulse $\hbar/\sigma$ (measured in eV) for $\epsilon=10^{-4}$. Pulse parameters were set such that $T=2.5\,\,\mathrm{ps}$ and $\boldsymbol{\mu}.\boldsymbol{v} = 0.041~\mathrm{eV.ps}$.  The pulse width axis has a log scale.}
    \label{fig:transient}
\end{figure}

\emph{Example 1:} Consider a qubit system(with the two levels denoted as $\{\ket{0},\ket{1}\}$) and initially in the state $\ket{0}$ that is acted upon by random unitary gates, drawn uniformly from the Haar measure, at uniform intervals of time. Constructing the time series of expectation values of the three appropriately defined Pauli operators $\{\hat{\sigma}_x,\hat{\sigma}_y,\hat{\sigma}_z\}$  on the qubit Hilbert space (besides the identity operator), we can compute $\misc$ corresponding to each of these time series. In our numerical experiment, the number of random unitaries acting on the qubit was set at 10000, and the entire experiment was repeated 84 times. For the tolerance parameter $\epsilon$ set to both $10^{-2}$ and $10^{-4}$, the resulting TD matrices (each of dimensions $5000\times5000)$, were always found to be full rank, yielding $\misc = \sqrt{5000} = 70.71$, indicating that the dynamics corresponding to this scenario are far more complicated than to be captured meaningfully by the indicated length of time series. Although this is an extreme example, it goes to show that the complexity witnessed, even in a manifestly unitary evolution of the qubit, can be arbitrarily large for a time-dependent interaction for time series constructed with respect to experimental time.   

\emph{Example 2:} Consider a two-level atom (with energy levels separated by $E = \hbar\omega_0$) interacting with a classical electromagnetic field. For simplicity, we assume that the carrier frequency is $\omega_0$ and field envelope of the pulse is Gaussian
\begin{equation}
    \bm{f}(t) = \frac{\eta e^{-(t-T)^2/2\sigma^2}e^{-i\omega_0 (t-T)}}{\sqrt{2\pi\sigma^2}} \bm{v},
\end{equation}
$\bm{v}$ is the field polarization vector, $T$ is the central pulse time, and $\eta,\sigma$ are pulse intensity and width (in time) respectively. 
If the atom is initially in its ground state, then the atom-field interaction results in the expectation value of the inversion operator~\cite{mandel1995optical} to be
\begin{equation}\label{eq:sigmaztransient}
    \langle\hat{\sigma}_z(t) \rangle = -\cos\left( \frac{2\boldsymbol{\mu}.\boldsymbol{v}}{\hbar}\left[ 1+ \mathrm{erf}\left(\frac{t-T}{\sqrt{2}\sigma}\right) \right] \right) 
\end{equation}
where $\boldsymbol{\mu}$ is the transition dipole moment of the atom, and $\mathrm{erf}(t)$ is the error function defined in the usual way. 

As the pulse width (in units of energy) varies from very narrow to very broad, there is a change in the nature of atomic dynamics and hence a corresponding change in the value of $\misc$ observed, as displayed in Fig. \ref{fig:transient}. For a very short pulse, the population oscillates sinusoidally (at the characteristic Rabi frequency) yielding a small value for $\misc$. As the pulse width increases, transient effects cause the $\misc$ to increase. However, this increase is not unbounded - for very broad pulses, the atomic dynamics are essentially a step function switch from an initial value of $-1$ to $-\cos\left(4\bm{\mu}.\bm{\varepsilon}/\hbar\right)$, and the finite resolution and length of time series employed fails to capture most dynamics and yields a constant, somewhat smaller number compared to the maximum $\misc$ observed as pulse width increases.   

\section{Analytical Lower Bound on $\misc$ for the Jeynes-Cummings (JC) Model}
\label{appendix:b}

To derive an analytical expression for a lower bound on $\misc$ for our model of JC atom interacting with a thermal mode, we begin with the following observation: tghe column vectors of the TD matrix $M^{\Upsilon_l(t)}$ corresponding to the exponential time series $\Upsilon_l(t) = e^{i\omega_l t}, \omega_l\in\mathbb{R}, t\in\mathbb{N}$, span a one-dimensional space, which in turn are disjoint for $\omega_l\neq \omega_m$. This implies that $\mathrm{rank}(M^{\Upsilon_l(t)}) = 1$ and $\mathrm{rank} (M^{\sum_{l=1}^{L}\Upsilon_l(t)}) = \sum_{l=1}^L \mathrm{rank}(M^{\Upsilon_l(t)}) = L$.

The time series corresponding to the inversion operator in Eq.~(\ref{eq:13}) is notated as the $(K+1)-$vector, $\bm{\langle\hat{\sigma}_z\rangle}\equiv \{\langle\hat{\sigma}_z(\lambda t)\rangle\},\, \lambda t\in\{0,1,\dots,K\}$. The above discussion immediately implies that each cosine term in the sum in Eq.~(\ref{eq:13}) generates a corresponding TD matrix of analytical rank $2$ (corresponding respectively to positive and negative frequency). The disjoint column spaces spanned by the orthogonal column vectors further imply that the ranks of the components are additive, from which we infer that analytical rank of the TD matrix {$M^{\bm{\langle\hat{\sigma}_z\rangle}}$ (and therefore $\mathrm{MISC}_0$) corresponding to Eq.~(\ref{eq:13})} diverges. 

To compute $\mathrm{MISC}_{\epsilon}$, we partition the signal time series as
\be
    \bm{\langle\hat{\sigma}_z\rangle} =  \bm{\langle\hat{\sigma}_z\rangle}_R +  \bm{\langle\hat{\sigma}_z\rangle}_{\mathrm{perturb}} ;\,\,\,\, \langle\hat{\sigma}_z(\lambda t)\rangle_R = \sum_{n=0}^{R}P_n \cos(\lambda t\sqrt{n+1}),\,\,\, \langle\hat{\sigma}_z(\lambda t)\rangle_{\mathrm{perturb}} = \sum_{n=R+1}^{\infty}P_n \cos(\lambda t\sqrt{n+1})
\ee
where $P_n = (\bar{n}/(1+\bar{n}))^n$. The TD matrix corresponding to the first term in the above decomposition $M^{\bm{\langle\hat{\sigma}_z\rangle}_R}$ is rank-deficient. In fact, $\mathrm{rank}(M^{\bm{\langle\hat{\sigma}_z\rangle}_R})=2R+1$ from the earlier arguments. To quantify the effect of $\bm{\langle\hat{\sigma}_z\rangle}_{\mathrm{perturb}}$ on the singular value spectrum of $M^{\bm{\langle\hat{\sigma}_z\rangle}_R}$ we invoke Weyl's inequality \cite{Stewart90perturbationtheory}, whereby
\begin{equation}
    |~\Sigma_j(M^{\bm{\langle\hat{\sigma}_z\rangle}}) - \Sigma_j(M^{\bm{\langle\hat{\sigma}_z\rangle}_R})~| \leq ||M^{\bm{\langle\hat{\sigma}_z\rangle}_{\mathrm{perturb}}}||_2, \,\,\,\,\, \forall\,\,\, j\in\{1,\dots,\lceil K/2 \rceil\},
\end{equation}
where $\Sigma_j(M)$ is the $j$-th singular value of the matrix $M$ (arranged in descending order of magnitude) and $||\,\cdot\,||_2$ denotes the spectral matrix norm, defined as 
\begin{equation}
    ||M||_2 \equiv \substack{\mathrm{max}\\||x||_2=1} \,\,||Mx||_2.
\end{equation}
The matrix norm of the TD matrix $M^{\bm{\langle\hat{\sigma}_z\rangle}_{\mathrm{perturb}}}$ is bounded by its Frobenius norm, which in turn is proportional to $\mathrm{exp}(-\beta\hbar\omega(R+1))$, allowing us to conclude that, asymptotically, the only effect of the component $\bm{\langle\hat{\sigma}_z\rangle}_{\mathrm{perturb}}$ will be to increase the rank of the total TD matrix $M^{\bm{\langle\hat{\sigma}_z\rangle}}$ without changing the $2R+1$ singular values or the singular vectors of the $\bm{\langle\hat{\sigma}_z\rangle}_R$ component. Thus, the component $\bm{\langle\hat{\sigma}_z\rangle}_R$ amounts to a recomposition of the time series with the $2R+1$ largest singular values. The root mean square perturbation (defined in Eq.~(\ref{eq:deltar})) is then 

\begin{equation}\label{eq:rms}
\Delta_R^{\mathrm{JC}} = \frac{|| \bm{\langle\hat{\sigma}_z\rangle} - \bm{\langle\hat{\sigma}_z\rangle}_R   ||_2} {||\bm{\langle\hat{\sigma}_z\rangle}||_2} = \frac{||\bm{\langle\hat{\sigma}_z\rangle}_{\mathrm{perturb}}||_2}{||\bm{\langle\hat{\sigma}_z\rangle}||_2}
\end{equation}
where $||\bm{v} ||_2 = \sqrt{ \sum_{\lambda t=0}^{K} |v(\lambda t)|^2 }$ is the vector $2$-norm. The numerator in Eq.~(\ref{eq:rms}) can be bounded, using the Cauchy-Schwartz inequality, as
\begin{equation}
     ||\bm{\langle\hat{\sigma}_z\rangle}_{\mathrm{perturb}}||_2 \leq \sqrt{K}~\frac{(1-e^{-\beta\hbar\omega})e^{-(R+1)\beta\hbar\omega}}{\sqrt{1-e^{-2\beta\hbar\omega}}}
\end{equation}
For the denominator, the reverse triangle inequality $||\bm{x}+\bm{y}||_2\geq|~ ||\bm{x}||_2-||\bm{y}||_2 ~|$ for vectors $\bm{x},\bm{y}$ gives:
\begin{equation}
   ||\bm{\langle\hat{\sigma}_z\rangle}||_2  \geq \sqrt{K} |1-2e^{-\beta\hbar\omega}|
\end{equation}
Putting these two together, we get the following upper bound on the magnitude of the RMS perturbation:
\begin{equation}
    \Delta_R^{\mathrm{JC}} \leq \frac{(1-e^{-\beta\hbar\omega})e^{-\beta\hbar\omega(R+1)}}{|1-2e^{-\beta\hbar\omega}|\sqrt{1-e^{-2\beta\hbar\omega}}}.
\end{equation}
Abbreviating $f = (1-e^{-\beta\hbar\omega})/(|1-2e^{-\beta\hbar\omega}|\sqrt{1-e^{-2\beta\hbar\omega}})$, we then get the following lower bound, using the definition of $\misc$ set out in Eq.~(\ref{eq:misce}):
\begin{equation}\label{eq:lbjc}
\misc ~\geq~ \sqrt{2\left\lfloor\frac{1}{\beta\hbar\omega}\ln\left(\frac{f}{\epsilon}\right)\right\rfloor + 1}.
\end{equation}
Note that the form of Eq.~(\ref{eq:lbjc}) clearly demonstrates that the $\misc$ is a monotonically decreasing function of the tolerance measure $\epsilon$. From Figure \ref{fig:jc}, we also see that the bound holds for the range of average photons per mode indicated in the figure for $\epsilon=10^{-2},10^{-3},\textrm{\,and\,} 10^{-4}$. 

\section{Details of Spin-Boson Dynamics Simulation}
\label{appendix:e}

The hierarchy of equations of motion (HEOM)~\cite{ishizaki2005quantum,ishizaki2020prerequisites,tanimura2020numerically} provides a nonperturbative framework to solve for the dynamics of open system strongly coupled to a non-Markovian bath at finite temperature. In this appendix, we first recap the basics of the HEOM formalism, before laying out the explicit equations that correspond to the Drude-Lorentz-like power spectral density of Eq.~(\ref{eq:sd}). 

In order to write down the path-integral form of the reduced density matrix for the two-level system (TLS), we first consider the fermionic coherent state $\ket{\psi}$, for which the equations $\hat{\psi}\ket{\psi}=\psi\ket{\psi}$ and $\bra{\psi}\hat{\psi}^{\dag} = \overline{\psi}\bra{\psi}$ hold. The Grassmann variables $\psi$ and $\overline{\psi}$ characterize the fermionic density matrix corresponding to the TLS state. 
The path-integral version of the TLS dynamics, for the factorizable initial state $\hat{\rho}(t_0) = \hat{\rho}_S(t_0)\otimes\hat{\rho}_{B}$ where $\hat{\rho}_S(t_0)$ is the initial TLS state and $\hat{\rho}_B$ is the canonical equilibrium state of the bath corresponding to temperature $T$, is
\begin{equation}\label{eq:pathintegral}
    \rho(\overline{\psi},\psi';t) = \int D\overline{\psi}D\psi\int D\overline{\psi}'D\psi'\rho(\overline{\psi}_0,\psi_0';t_0) \,\,e^{iS[\overline{\psi},\psi]/\hbar} F_{FV}[\overline{\psi},\psi;\overline{\psi}',\psi'] e^{-iS[\overline{\psi}',\psi']/\hbar}
\end{equation}
where $S[\overline{\psi},\psi]$ is the action $S = \int dt L(\overline{\psi},\psi)$ corresponding to the TLS Lagrangian $L(\overline{\psi},\psi)$:

\begin{equation}
L(\overline{\psi},\psi) = \frac{i\hbar}{2}(\overline{\psi}\dot{\psi} - \overline{\dot{\psi}}\psi) - \frac{\hbar\omega_0}{2}(\overline{\psi}\psi - \psi\overline{\psi})
\end{equation}
and the Feynman-Vernon influence functional~\cite{feynman1963theory} $F_{FV}$ is
\begin{equation}\label{fvdef}
    F_{FV}(\overline{\psi},\psi;\overline{\psi'},\psi') = \mathrm{exp}\left(-\frac{1}{\hbar}\int_{t_0}^{t}ds\int_{t_0}^{s} W^{\times}(s)\times(D_1(s-s')W^{\times}(s')-\frac{i\hbar}{2}D_2(s-s')W^{\circ}(s'))\right)
\end{equation}
where $D_1(t) =  (\hbar/\pi) \int_0^{\infty} d\omega J(\omega) \coth(\beta\hbar\omega/2)\cos(\omega t)$ is the symmetrized correlation function, and $D_2(t) = (2/\hbar)\int_0^{\infty}d\omega J(\omega)\sin \omega t$ is the bath response function, and 
\begin{equation}
    W^{\times}(t) = W(\overline{\psi}(t),\psi(t)) - W(\overline{\psi'}(t),\psi'(t)), ~  W^{\circ}(t) = W(\overline{\psi}(t),\psi(t)) + W(\overline{\psi'}(t),\psi'(t)).
\end{equation}
The HEOM framework, as applied to Lorentzian spectral densities, hinges on a factorization of the influence functional $F_{FV}$, which for the spectral density of Eq.~(\ref{eq:sd}) takes the form
\begin{equation}
    F_{FV} = \mathrm{exp}\left(\int_{t_0}^t ds\int_{t_0}^s ds'~ \Phi(s)[\Theta_1(s')+\Theta_2(s')]~ \mathrm{Re}(\gamma_+ e^{-\gamma_+ (s-s')}) \right)\times \prod_{m=1}^{\infty} \mathrm{exp}\left( \int_{t_0}^{t} ds \int_{t_0}^s ds'~ \Phi(s) \Psi_m(s')~ \nu_m e^{-\nu_m(s-s')}\right)
\end{equation}
where $\nu_m = 2\pi m/\beta\hbar$ is the $m$-th Matsubara frequency and the various Grassmann variable-valued functions are defined as
\begin{align}
    &\Phi(t) = iW^{\times}(t)\noindent\nonumber\\
    &\Theta_1(t) = \frac{i\zeta}{4\beta\hbar\epsilon}\left(\frac{\gamma}{\omega_0}\right)\left[\mathrm{Re}(c_1)W^{\times}(t) - i\mathrm{Re}(c_2)W^{\circ}(t)\right],~ c_1 = i\beta\hbar\gamma_- \cot\left(\frac{\beta\hbar\gamma_+}{2}\right)   \nonumber\noindent\\
    &\Theta_2(t) = \frac{i\zeta}{4\beta\hbar\epsilon}\left(\frac{\gamma}{\omega_0}\right)\left[\mathrm{Im}(c_1)W^{\times}(t) - i\mathrm{Im}(c_2)W^{\circ}(t)\right], ~  c_2 = i\beta\hbar\gamma_- \nonumber\noindent\\
    & \Psi_m(t) = \frac{i\zeta}{\beta\hbar\omega_0}~\frac{2\gamma^2\gamma_+\gamma_-}{(\nu_m^2-\gamma_+^2)(\gamma_-^2-\nu_m^2)}~ W^{\times}(t).
\end{align}
Choosing the Matsubara index $M$ such that $\nu_M \geq \omega_0$ holds, the Feynman-Vernon influence functional reduces to
\begin{align}\label{fvequation}
    F_{FV} &= \mathrm{exp}\left(\int_{t_0}^t ds\int_{t_0}^s ds'~ \Phi(s)[\Theta_1(s')+\Theta_2(s')]~ \mathrm{Re}(\gamma_+ e^{-\gamma_+ (s-s')}) \right) \times \prod_{m=1}^{M} \mathrm{exp}\left( \int_{t_0}^{t} ds \int_{t_0}^s ds'~ \Phi(s) \Psi_m(s')~ \nu_m e^{-\nu_m(s-s')}\right) \nonumber\noindent\\
    &~~\times~~\prod_{m=M+1}^{\infty} \mathrm{exp}\left( \int_{t_0}^{t} ds ~ \Phi(s)\Psi_m(s) \right),
\end{align}
where we have used the simplification, $\int_{0}^{\infty} dt\, \nu_m e^{-\nu_m t}\, f(t) \approx f(0), \, \, m \geq M$, for any well-behaved function $f(t)$, to replace the factor $\nu_m e^{-\nu_m(s-s')}$ by the Dirac delta function $\delta(s-s')$ under the integral in Eq.~(\ref{fvequation}).
Introducing the family of auxiliary density functions
\begin{align}\label{eq:adms}
    \rho^{(n)}_{j_1,\dots,j_M}(\overline{\psi},\psi';t) &= \int D\overline{\psi}D\psi\int D\overline{\psi}'D\psi'\rho(\overline{\psi}_0,\psi_0';t_0)\left\{-\int_{t_0}^{t}ds\hat{\Theta}_1(s)\mathrm{Re}[\gamma_+ e^{-\gamma_+ (t-s)}]\right\}^n  \left\{-\int_{t_0}^{t}\hat{\Theta}_2(s)\mathrm{Im}[\gamma_+ e^{-\gamma_+ (t-s)}]\right\}^n  \noindent\nonumber\\
    &\times \prod_{m=1}^M \left\{ -\int_{t_0}^{t}ds\hat{\Psi}_m(s)\nu_m e^{-\nu_m(t-s)}\right\}^{j_m}  e^{iS[\overline{\psi},\psi]/\hbar} F_{FV}[\overline{\psi},\psi;\overline{\psi}',\psi'] e^{-iS[\overline{\psi},\psi]/\hbar} ,
    \end{align}
the corresponding hierarchy of equations of motion take the final form
\begin{align}\label{eq:hierarchy}
       \frac{\partial \rho^{(n)}_{j_1,\dots,j_M}(t) }{\partial t} &= -i \! \! \left[ \hat{\mathcal{L}} + 2n\gamma + \sum_{m=1}^{M}(j_m\nu_m + \hat{\Phi}\hat{\Psi}_m) + \hat{\Xi} \right] \!\!
            	\hat{\rho}^{(n)}_{j_1,\dots,j_M} 
            - \hat{\Phi}\hat{\rho}^{(n+1)}_{j_1,\dots,j_M} - n\gamma(\hat{\Theta}_1+\hat{\Theta}_2)\hat{\rho}^{(n-1)}_{j_1,\dots,j_M} - \sum_{m=1}^{M}\hat{\Phi}\hat{\rho}^{(n)}_{j_1,\dots,j_m+1,\dots,j_M}  \nonumber\noindent\\
            &- \sum_{m=1}^{M}j_m\nu_m\hat{\Psi_m}\hat{\rho}^{(n)}_{j_1,\dots,j_{m-1},\dots,j_M}.
\end{align}
Note that the above system of equations is expressed in operator form, as opposed to as functions of the Grassmann variables. This is achieved via the substitutions $W^{\times}(t)\rightarrow\hat{W}^{\times}$ and $W^{\circ}(t)\rightarrow\hat{W}^{\circ}$, where the respective actions of the superoperators are defined as  $\hat{W}^{\times}\hat{f} = \hat{W}\hat{f} - \hat{f}\hat{W}$ and $\hat{W}^{\circ}\hat{f} = \hat{W}\hat{f} + \hat{f}\hat{W}$, where $\hat{f}$ is any operator. Further, $\hat{\mathcal{L}}$ is the Liouvillian of the TLS $\hat{\mathcal{L}}\hat{\rho} = [\hat{H}_0,\hat{\rho}]$ where $\hat{H}_0$ is the TLS Hamiltonian. Finally, the renormalization operator $\hat{\Xi}$ can be evaluated as
\begin{equation}
    \hat{\Xi} = \sum_{m=1}^{\infty} \hat{\Phi}\hat{\Psi}_m = \frac{i\zeta}{\beta\hbar\omega_0}\frac{\gamma^2}{\gamma_+\gamma_-}\left[ 1 - \left( \frac{\gamma_-^2}{\gamma_-^2 - \gamma_+^2}\left(\frac{\beta\hbar\gamma_+}{2}\right)\cot\left(\frac{\beta\hbar\gamma_+}{2}\right) -  \frac{\gamma_+^2}{\gamma_-^2 - \gamma_+^2}\left(\frac{\beta\hbar\gamma_-}{2}\right)\cot\left(\frac{\beta\hbar\gamma_-}{2}\right)    \right)      \right] \hat{W}^{\times}\hat{W}^{\times}
\end{equation}
In the hierarchy of auxiliary density matrices defined in Eq.~(\ref{eq:adms}), the lowest member element, $\hat{\rho}^{(0)}_{0,\dots,0}$, is physically meaningful and is equal to the density matrix of the open system $\rho(t)$, as can be inferred from Eq.~(\ref{eq:pathintegral}). The higher order hierarchy elements encode in turn higher order non-Markovian effects that are propagated to the system density matrix through the system of coupled linear differential equations in Eq.~(\ref{eq:hierarchy}). In principle, solving the infinite hierarchy yields a non-perturbative solution of open system dynamics.

\begin{figure}
    \centering
    \includegraphics[width=1.0\linewidth]{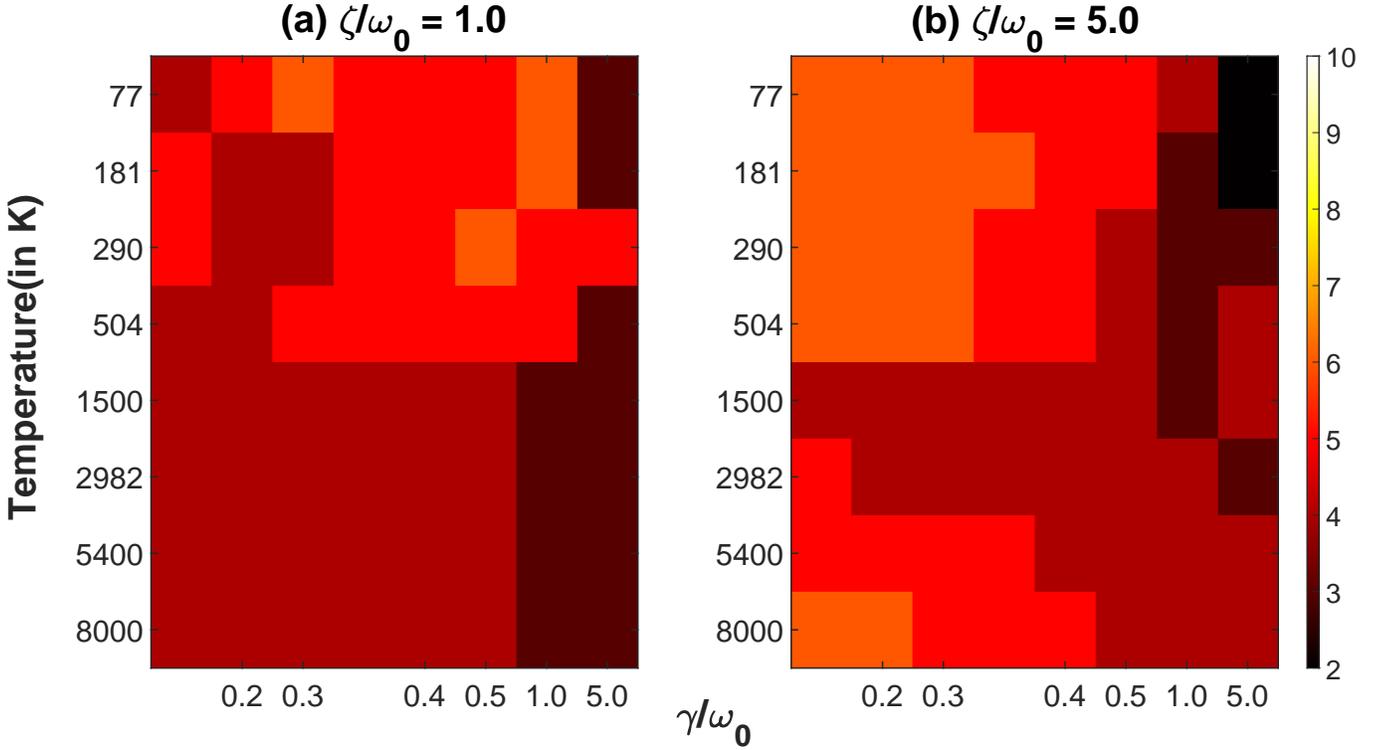}
    \caption{ Heat maps of simulation complexity $\misc$ calculated using expectation value of system observables $\{\hat{\sigma}_x,\hat{\sigma}_y,\hat{\sigma}_z\}$ describing energy dissipation and coherence dynamics in the spin-boson model. The depth of hierarchy D is set to 8, the number of Matsubara frequencies M are set to 5 for all points on grid, and $\epsilon=10^{-4}$. The hybrid process interaction matrix is obtained by setting $W_1=W_2=1/2$. (a) $\zeta/\omega_0 = 1.0$ corresponding to weak coupling between TLS and bosonic environment. (b) $\zeta/\omega_0 = 5.0$ corresponding to strong coupling between TLS and bosonic environment.    }  
    \label{fig:sbheatmixed}
\end{figure}

In practice, however, one can meaningfully terminate the hierarchy to obtain a finite system of differential equations, that correspond to a numerically exact solution. For indices $n,j_1,\dots,j_M$ for which
\begin{equation}
    D \equiv n + \sum_{m=1}^M j_m \,\gg\, \frac{\omega_0}{\mathrm{min}(\gamma,\nu_1)}, 
\end{equation}
the HEOM equation Eq.~(\ref{eq:hierarchy}) reduces to the termination relation~\cite{ishizaki2010quantum}
\begin{equation}\label{eq:terminator}
    \frac{\partial}{\partial t} \hat{\rho}^{n}_{j_1,\dots,\j_M} = -\left( i\hat{\mathcal{L}} + \sum_{m=1}^{M}\hat{\Phi}\hat{\Psi}_m + \hat{\Xi} \right)  \hat{\rho}^{n}_{j_1,\dots,\j_M}.
\end{equation}
The mixed index $D$ is the designated depth of hierarchy. 

Our numerical solutions assume a constant number of Matsubara frequencies $M=5$ and depth of hierarchy $D=8$. These numbers were picked so as the terminators in Eq.~(\ref{eq:terminator}) are valid for the entire parameter regime of $\{1/T,\zeta,\gamma\}$ studied in the heat maps in Fig. \ref{fig:sbheat} in the main text. Note that the orders of hierarchical equations are not optimal for all points on the grids -- for instance, the points corresponding to high temperature and/or fast dynamics on the grid do not need as many Matsubara frequencies for a convergent solution as was chosen. The $ (D,M) $ pair was, however, fixed for all calculations to facilitate a consistent comparison of simulation complexity for different values of parameters characterising the system-environment interaction. The convergence of solutions was evaluated by verifying that the dynamics resulting from neighbouring  $ (D,M) $ pairs was close to that corresponding to $(D=8,M=5) $, and it was found that steady state was achieved for the chosen values for all points on the grid.

In the heat maps of Fig. \ref{fig:sbheat} 1a and 1b, in order to induce only population dynamics and no dephasing associated with the $T_1$ process ($W_1=1, W_2 = 0$), the initial state of the two-level system is set to 
$\hat{\rho}(0) = \ket{1}\bra{1}\otimes\hat{\rho}_B^{eq},$ where $\hat{\rho}_B^{eq}$ is the thermal equilibrium state of the bath at temperature $T$. To describe the pure dephasing of the $T_2$ type (heat maps of Fig.~\ref{fig:sbheat} 2a and 2b) for which $W_1=0, W_2=1$, we instead set the initial system-bath state to $ \hat{\rho}(0) = 0.5\,(\ket{0}+\ket{1})(\bra{0}+\bra{1})\otimes\hat{\rho}_B^{eq}$. The dynamics of the corresponding operator expectations are displayed in Figs. \ref{fig:signalplot1}-\ref{fig:signalplot4}, for both the $T_1$ and $T_2$ processes (weak and strong coupling) for the range of system-bath parameters that appear on the grids in Fig. \ref{fig:sbheat}. 

Finally, we also compute the $\misc$ number for the case when both the $T_1$ and $T_2$ processes operate simultaneously in the same system, for instance if $W_1=W_2=1/2$. The initial state in this case is set to $ \hat{\rho}(0) = 0.5\,(\ket{0}+\ket{1})(\bra{0}+\bra{1})\otimes\hat{\rho}_B^{eq}$. Results are displayed as heat maps for same parameter intervals of inverse correlation time ($1/\gamma$) and coupling strength ($\zeta$) as studied in Fig.~\ref{fig:sbheatmixed}. The $\misc$ for the hybrid process increases less strongly with temperature than for the $T_1$ process, while all the other trends in $\misc$ expectedly hold (increase with strength of coupling $\zeta$ and inverse correlation time $1/\gamma$).

\begin{figure}[p]
    \centering
    \includegraphics[width=1.0\linewidth]{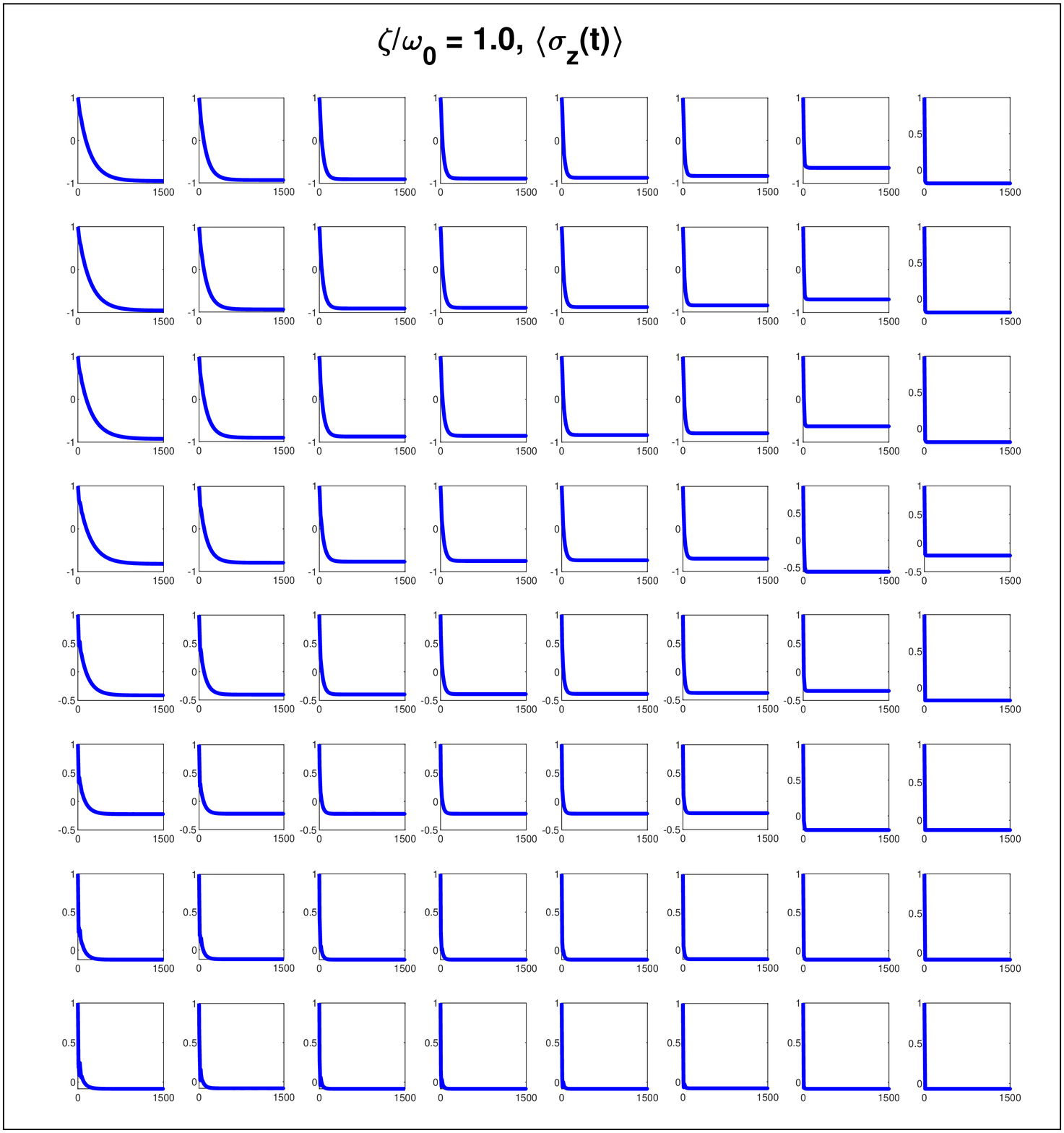}
    \caption{Inelastic dissipation dynamics of the population inversion operator $\langle\hat{\sigma_z}(t)\rangle$ corresponding to small system-bath coupling $\zeta/\omega_0=1.0$, plotted against time (in natural units). Each panel in the $8\times 8$ matrix figure corresponds to its respective point on the grid of Fig. \ref{fig:sbheat} 1(a). The time series was constructed using 30,000 values of $\langle\hat{\sigma}_z(t)\rangle$, uniformly sampled between normalised times 0 and 1500 for all points on the grid. The tolerance parameter was set to $\epsilon=10^{-4}$.}
    \label{fig:signalplot1}
\end{figure}

\begin{figure}[p]
    \centering
    \includegraphics[width=1.0\linewidth]{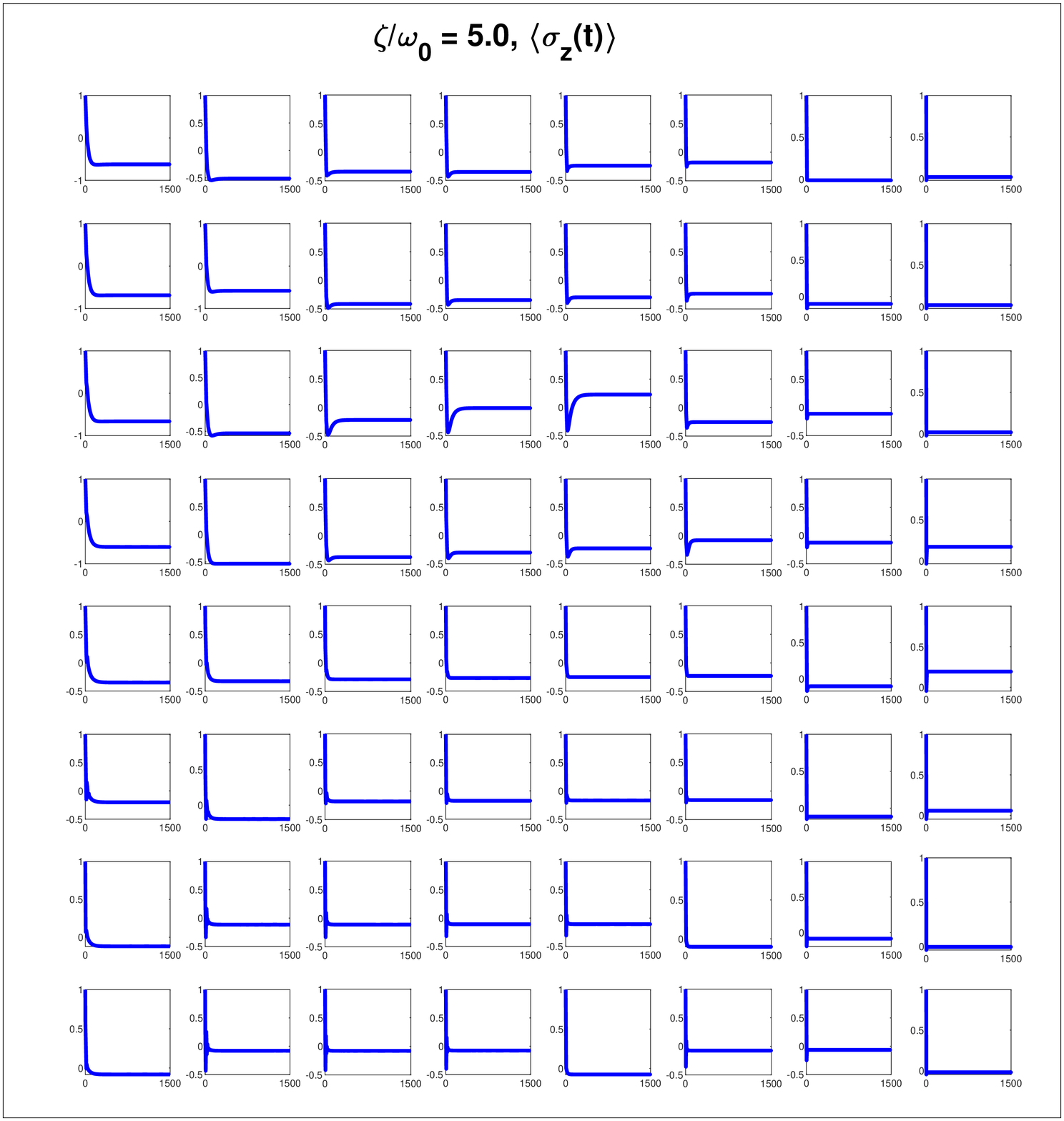}
    \caption{Inelastic dissipation dynamics of the population inversion operator $\langle\hat{\sigma_z}(t)\rangle$ corresponding to strong system-bath coupling $\zeta/\omega_0=5.0$, plotted against time (in natural units). Each panel in the $8\times 8$ matrix figure corresponds to its respective point on the grid of Fig. \ref{fig:sbheat} 1(b). The time series was constructed using 30,000 values of $\langle\hat{\sigma}_z(t)\rangle$, uniformly sampled between normalised times 0 and 1500 for all points on the grid. The tolerance parameter was set to $\epsilon=10^{-4}$.}
     \label{fig:signalplot2}
\end{figure}

\begin{figure}[p]
    \centering
    \includegraphics[width=1.0\linewidth]{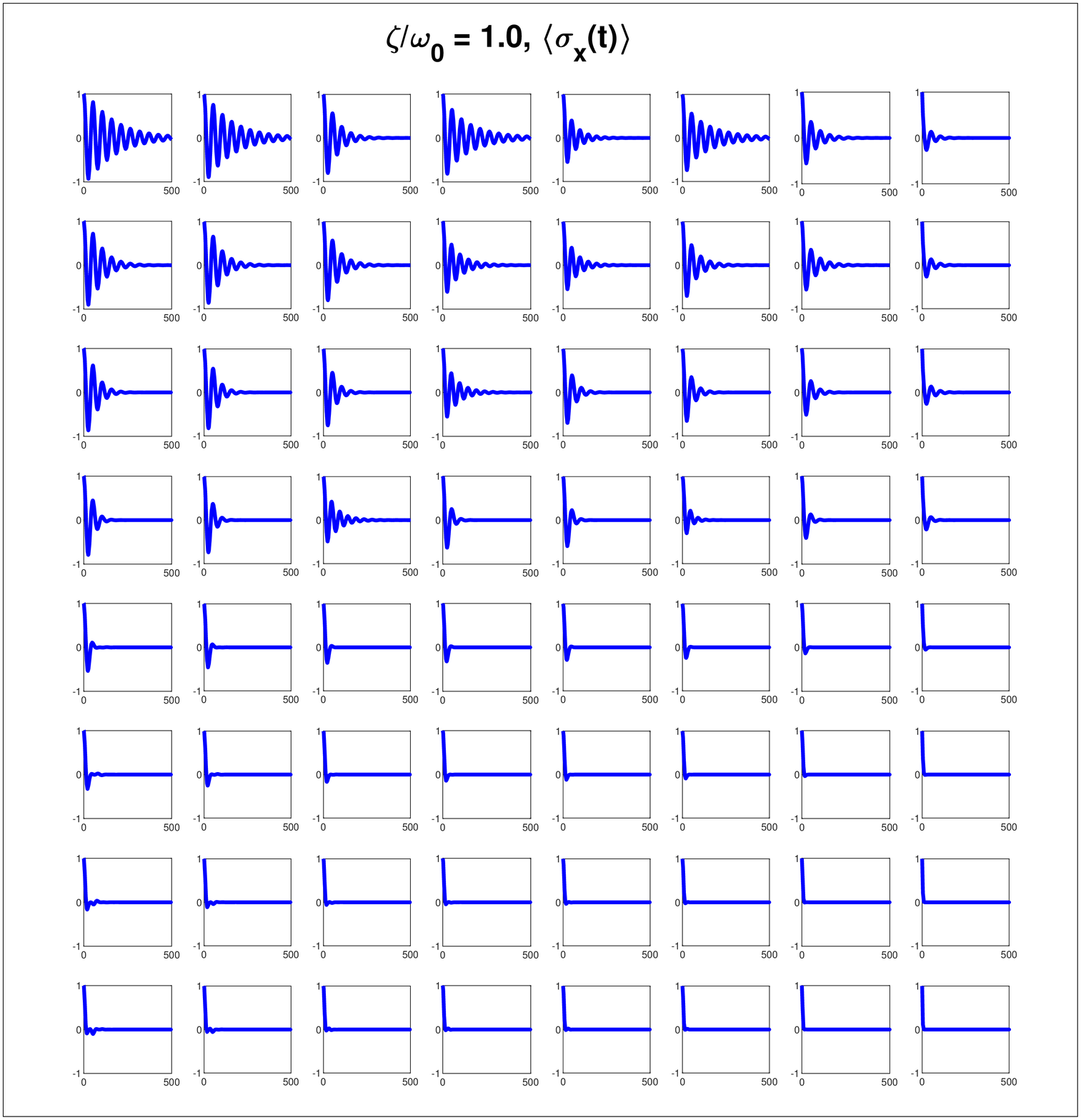}
    \caption{Elastic dephasing relaxation of the $\langle\hat{\sigma}_x(t)\rangle$ coherence observable corresponding to small system-bath coupling $\zeta/\omega_0=1.0$, plotted against time (in natural units). Each panel in the $8\times 8$ matrix figure corresponds to its respective point on the grid of Fig. \ref{fig:sbheat} (2)(a). The time series was constructed using 29,441 values of $\langle\hat{\sigma_x}(t)\rangle$, uniformly sampled between normalised times 0 and 500 for all points on the grid. The tolerance parameter was set to $\epsilon=10^{-4}$.}
     \label{fig:signalplot3}
\end{figure}

\begin{figure}[p]
    \centering
    \includegraphics[width=1.0\linewidth]{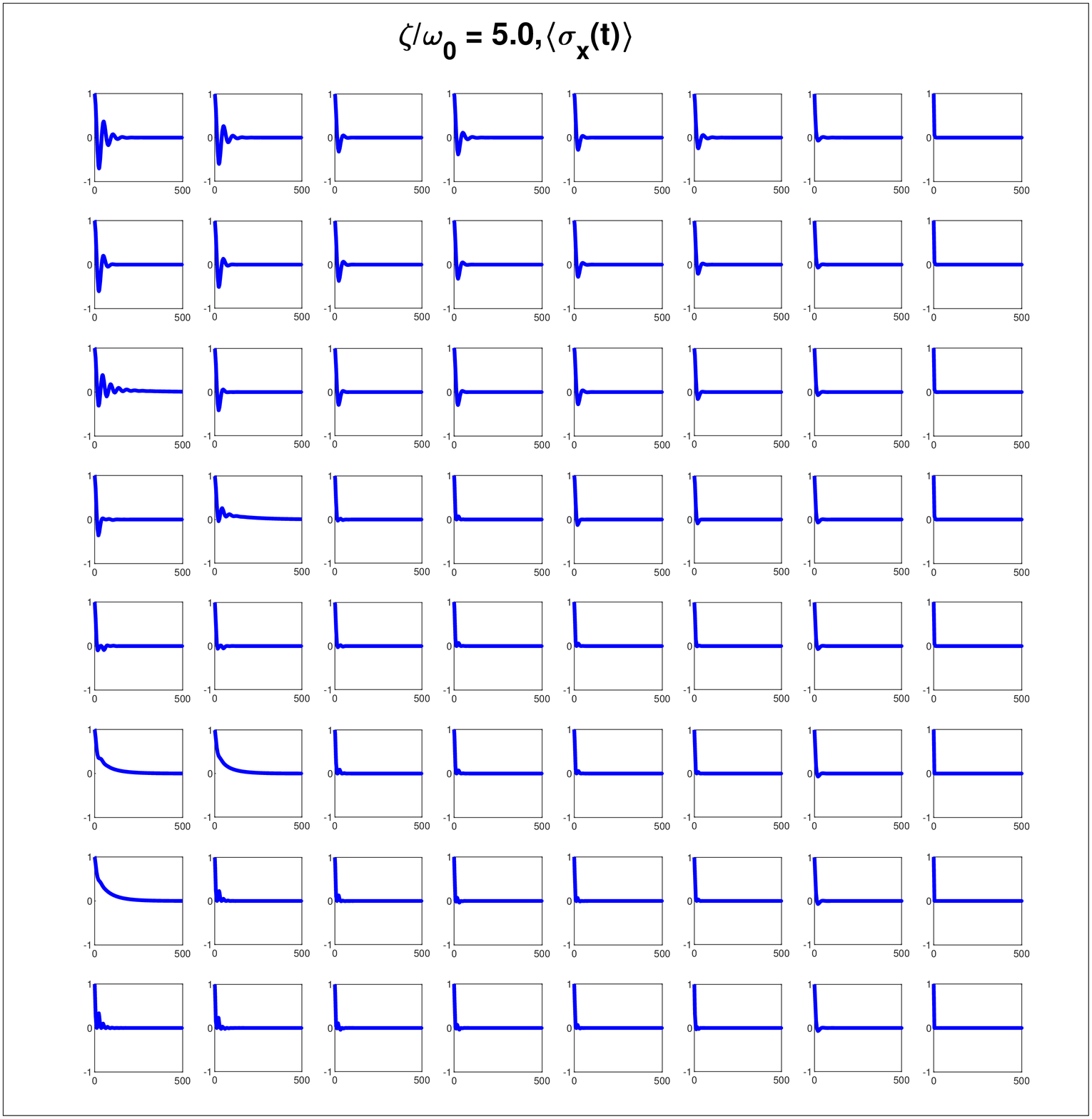}
    \caption{Elastic dephasing relaxation of the $\langle\hat{\sigma}_x(t)\rangle$ coherence observable corresponding to strong system-bath coupling $\zeta/\omega_0=5.0$, plotted against time (in natural units). Each panel in the $8\times 8$ matrix figure corresponds to its respective point on the grid of Fig. \ref{fig:sbheat} (2)(b). The time series was constructed using 29,441 values of $\langle\hat{\sigma_z}(t)\rangle$, uniformly sampled between normalised times 0 and 500 for all points on the grid. The tolerance parameter was set to $\epsilon=10^{-4}$.}
     \label{fig:signalplot4}
\end{figure}

\section{Proposed Pump-Probe Experiment and its Numerical Simulation}\label{appendix:c}

The spectroscopic arrangement considered here involves a non-collinear pump-probe ($PP'$) setup where the sample is illuminated first by the pump pulse, travelling in the $\bm{k}_P$ direction, followed by the probe pulse travelling in $\bm{k}_{P'}$ direction. The experimental arrangement was originally proposed for performing quantum process tomography (QPT) for EET in photosynthetic light harvesting complexes~\cite{yuen2014ultrafast}. It was later adapted to witness uniquely quantum coherence in EET transfer process in photosynthetic light harvesting complexes~\cite{marcus2019towards}.
 
The spectroscopic signal is defined as the differential loss of intensity of the probe pulse with and without the action of the pump pulse.
This can be most straightforwardly calculated by averaging over the responses of individual molecules in the sample, which in turn are evaluated by tracking the populations of the various electronic levels of each molecule in the sample. We accomplish this by numerically integrating the time-dependent Schr\"{o}dinger equation set out by the sum of time-independent Frenkel-Holstein Hamiltonian (defined in the main text) and the time-dependent dipole-field interaction of the molecule at position $\bm{r}$ interacting with a succession of pulses given by
\be
    \hat{H}_I(t) = -\bm{\mu}.\left(\sum_{n=P,P'}[\vartheta_n(t-t_n)\bm{\hat{e}}_n e^{i\bm{k}_n.\bm{r} + \phi_n} + c.c] \right) = \sum_{n=P,P'} \hat{H}_I^{n,+}(t)e^{i\bm{k}_n.\bm{r} + \phi_n} + \hat{H}_I^{n,-}(t)e^{-i\bm{k}_n.\bm{r} - \phi_n}
\label{eq:hlm}
\ee
where $\vartheta_n(t)$ is the wavepacket envelope, $t_n$ is the central time of the $n$-th pulse, and $\phi(n,\bm{r}) = \bm{k}_n.\bm{r} + \phi_n$ is the position-dependent phase of light-matter interaction corresponding to the $n$-th pulse. Further, we assume that the Condon approximation holds, allowing us to drop the position dependence of the dipole operator $\bm{\mu}$.

For the simulation, we assume that the pulse envelope is Gaussian,
\begin{equation}\label{eq:pulse}
    \vartheta_n(t) = \frac{\eta e^{-t^2/2\sigma^2}e^{-i\omega_n t}}{\sqrt{2\pi\sigma_n^2}}
\end{equation}
where $\omega_n$ is the central frequency of the pulse in the Fourier space, $\eta$ is the intensity of the pulse, and $\sigma_n$ is the width of the pulse in the time domain. We study resonant interactions of the coupled dimer by setting the central frequencies of the pump and probe pulses equal to the average energy gaps between the ground state manifold (GSM) and the two excitonic states of the singly-excited manifold (SEM)~\cite{marcus2019towards}, meaning there are four distinct $PP'$ experiments considered: (-,-) corresponds to both pump and probe pulses resonant with the lower excited state of the SEM, (-,+) corresponds to the pump pulse resonant with the lower state and the probe pulse resonant with the upper state, and so on.

A complete simulation of each $PP'$  experiment would involve aggregating over differential signals accumulated from all molecules of the ensemble, each interacting with the light pulses slightly differently depending on $\bm{r}$ in Eq.~(\ref{eq:hlm}). Instead, we show that this aggregation of individual responses can be accomplished with significantly lesser computational effort. 

We first borrow the notation and corresponding definitions from Ref.~\cite{yuen2014ultrafast}, and set $\phi(P,\bm{r})=0$ as only the relative phases of the pulses matter in the final expression for the signal. We then express the differential single molecule $PP'$  signal as the following overlap of perturbative wave functions of the coupled dimer, where the loss of intensity of the light pulse interacting with a single molecule is equated to the increase in population of the doubly-excited manifold (DSM) less the population of GSM of the coupled dimer as
\begin{align}
S_{PP'}[t,\phi(P',\bm{r})] &= - 2Re\left[(\bra{\psi_0(t)}\ket{\psi_{P-P'}(t)} + \bra{\psi_{P-P}(t)}\ket{\psi_{P-P'}(t)} + \bra{\psi_{P'-P'}(t)}\ket{\psi_{P-P'}(t)} )e^{-i\phi(P',\bm{r})}\right]\nonumber\noindent\\
&-2Re\bra{\psi_{P-P}(t)}\ket{\psi_{P'-P'}(t)} - \bra{\psi_{P-P'}(t)}\ket{\psi_{P-P'}(t)} +  \bra{\psi_{P+P'}(t)}\ket{\psi_{P+P'}(t)} \noindent\\
&+ 2Re\left[\bra{\psi_{P+P}(t)}\ket{\psi_{P'+P'}(t)}e^{i2\phi(P',\bm{r})}  + (\bra{\psi_{P+P'}(t)}\ket{\psi_{P'+P'}(t)}
+\bra{\psi_{P+P}(t)}\ket{\psi_{P+P'}(t)})e^{i\phi(P',\bm{r})} \right],  \nonumber
\end{align}
where the perturbative wavefunctions are defined as
\ben
    \ket{\psi_{n_1\pm \dots\pm n_{m}}(t)} &= & i^m\int_{t_0}^{t} dt_1\dots\int_{t_{m-1}}^{t_{m}}dt_m e^{-iH_0(t-t_1)} \left\{-\hat{H}_I^{n_1,\pm}(t_1)\right\}e^{-iH_0(t_1-t_2)}\dots e^{-iH_0(t_{m-1}-t_{m})}\nonumber\noindent\\
    &&\times \left \{-\hat{H}_I^{n_m,\pm}(t_m) \right\}e^{-iH_0 t_m} \ket{\psi_0}.
\een
The signal measured at the end of the experiment ($t \gg t_{P'}$)  is independent of time and can be expressed in terms of asymptotic wavefunctions, defined as $\ket{\psi} = \lim_{t\to\infty} e^{iH_0 t}\ket{\psi(t)}$.  Setting $\phi(P',\bm{r}) \longrightarrow \phi(P',\bm{r}) + \pi$, we average over two instances of the phase-dependent terms to get
\begin{align}
    S_{PP'} &= ~\lim_{t\to\infty}\frac{1}{2}( S_{PP'}[t,\phi(P',\bm{r})] + 
    S_{PP'}[t,\phi(P',\bm{r}) + \pi]) \nonumber\noindent\\
    &=\bra{\psi_{P+P'}}\ket{\psi_{P+P'}}   - \bra{\psi_{P-P'}}\ket{\psi_{P-P'}} - 2Re\bra{\psi_{P-P}} \ket{\psi_{P'-P'}},
\end{align}
where we have also dropped the term $\bra{\psi_{P+P}}\ket{\psi_{P'+P'}}e^{i 2\phi(P',\bm{r})}$,  which is proportional to exp($-\omega^2_{\beta\alpha}(\sigma_P^2+\sigma_{P'}^2)/2$), $\omega_{\beta\alpha}$ being the energetic difference between the two excitonic states of the SEM, and hence is vanishingly small in magnitude compared to the other terms. Furthermore,
\ben
    \bra{\psi_{P'-P'}}\ket{\psi_{P-P}} &=& (+i)\int_{t_0}^{t} dt_1 \bra{\psi_{+P'}(t_1)}\{-\hat{H}_I^{P'+}(t_1)\} e^{iH_0(t-t_1)}\ket{\psi_{P'-P'}(t)} \noindent\nonumber\\
    &=& i\bra{\psi_{+P}(t)}\int_{t_0}^{t}dt_1 e^{-iH_0(t-t_1)}\{-\hat{H}_I^{P'+}(t_1)\}\ket{\psi_{P-P}(t_1)}\nonumber\noindent\\
    &= &- \bra{\psi_{+P}}\ket{\psi_{P-P+P'}}.
\een
The second step is valid if we assume that the Gaussian wavepacket corresponding to the first pulse is sharp ($\sigma_P \ll t_P$), leading to
\be
    S_{PP'} = -\bra{\psi_{P-P'}}\ket{\psi_{P-P'}} + 2Re(\bra{\psi_{P'}}\ket{\psi_{P-P+P'}}) + \bra{\psi_{P+P'}}\ket{\psi_{P+P'}},
\ee
which is also the signal obtained by averaging over all the molecules of the ensemble~\cite{yuen2014ultrafast}. The three terms in the signal correspond to the three dominant processes that occur between ground and excited states of the coupled dimer, and $S_{PP'}$ is now the aggregate differential signal that is measured in the pump-probe spectroscopic setup. Note that this is calculated by averaging over only two single molecule responses whose phases are separated by $\pi$, instead of aggregating over all the molecules in the ensemble, as the theoretical recipe lays out. 

The specific parameters used in the simulation of the $PP'$ experiment, including those of the Frenkel-Holstein Hamiltonian (Eqs.~(\ref{eq:fh1}-\ref{eq:fh3}) in the main text) are listed in Table \ref{tab:PPparam}. The first-order differential equations corresponding to the time-dependent Schrodinger equations are solved numerically using the variable-coefficient differential equation ZVODE solver, detailed in \cite{brown1989vode}.

\begin{table}[h!]
    \centering
    \begin{tabular}{  p{3.4cm}  p{2.0cm}  }
    \hline
         Parameter(unit)  & Value  \\
         \hline
        $ \varepsilon_1(\mathrm{cm}^{-1})$\,\,\,\,\,\,  &15 3000 \\
      $  \varepsilon_2(\mathrm{cm}^{-1}) $ \,\,\,\,\,\, & 16 200\\
    $    \kappa(\mathrm{cm}^{-1}) $ \,\,\,\,\,\, & -162 \\
    $     \omega_1(\mathrm{cm}^{-1}) $ \,\,\,\,\,\, & 800\\
      $   \omega_2(\mathrm{cm}^{-1}) $\,\,\,\,\,\, & 1500   \\
   $     g_1 $\,\,\,\,\,\,& 0.1 \\
     $    g_2 $\,\,\,\,\,\,& 0.15 \\
     $    \sigma_p(\mathrm{cm}^{-1}/\mathrm{fs})$ \,\,\,\,\,\,& 322/103\\
      $   \eta_p(\mathrm{eV ps}/\mathrm{D}) $ \,\,\,\,\,\,& $5\times 10^{-4}$\\
         \hline
    \end{tabular}
    \caption{System and bath parameters setting out the Frenkel-Holstein Hamiltonian in Eqs.~(\ref{eq:fh1}-\ref{eq:fh3}) corresponding to the coupled dimer of allophycocyanin~\cite{womick2009exciton}, as well as parameters corresponding to the pump and probe pulses set out in Eq.~(\ref{eq:pulse}).} 
    \label{tab:PPparam}
\end{table}

\section{$\misc$ for DC pulse sequence}
\label{appendix:dcmisc}

Fig.~\ref{fig:dcesp} shows $\misc$ for $\epsilon = 0.1, 0.3, 0,5$ for the DC data. 
Larger values of $\epsilon$ gives lower complexity as expected and also shown in Fig.~\ref{fig:jc} and Table~\ref{tab:PPsim}.
More importantly, it shows that larger $\epsilon $ values shrink regions of apparent high complexity in the DC data.
The high complexity regions in parts of the heat map for small $\epsilon$ are, in fact, a reflection of the low SNR in the data for those wavenumbers.
Evaluating $\misc$ using varying $\epsilon$ for a data set can thus reveal its noise floor in a manner independent of the experimental setup.

\begin{figure}[h!]
    \centering
    \includegraphics[width=1.0\linewidth]{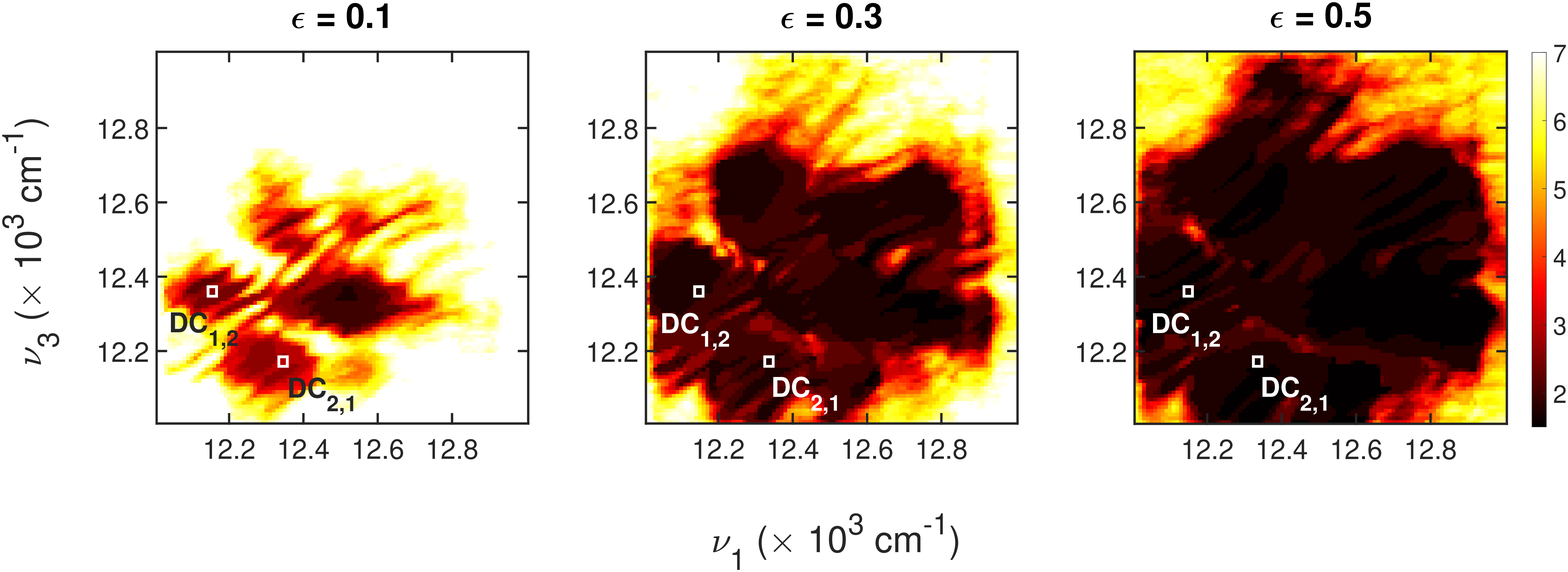}
    \caption{Heat maps of $\misc$ for DC pulse sequences over the $100\times 100$ grid of ($\nu_1,\nu_3$) points. Each point is generated from the time series of the complete rephasing signal $\tilde{E}^{(3)}(\nu_1,t_2,\nu_3)$, where $\nu_1$ and $\nu_3$ are proportional to the excitation and detection energies, and the complex emitted signal field is recorded over 2.9 ps for DC sequences, each sampled every 20 fs.  The cross-peaks for both pulse sequences are marked with white boxes on the map}
    \label{fig:dcesp}
\end{figure}

\end{widetext}

\end{document}